%% file: main.tex
\documentclass[usenatbib,fleqn]{mnras}
\pdfoutput=1 %for arxiv
\usepackage{amssymb}
\usepackage{amsmath}
\usepackage[dvipsnames]{xcolor}  
\usepackage{graphicx}
\usepackage[T1]{fontenc}
\usepackage{newtxtext,newtxmath}
\usepackage[figure,figure*]{hypcap}
\usepackage{soul}
\input{hyperlink-year-only-natbib-patch}
\input{villarrealzentner_macros}

\title[The Immitigable Nature of Assembly Bias]{The Immitigable Nature of Assembly
Bias: The Impact of Halo Definition on Assembly Bias}
\author[A.~S.~Villarreal et al.]{%
Antonio~S.~Villarreal$^{1}$\thanks{E-mail: \email{asv13@pitt.edu}},
Andrew~R.~Zentner$^{1}$\thanks{E-mail: \email{zentner@pitt.edu}},
Yao-Yuan~Mao$^{1}$, %\thanks{E-mail: yymao@pitt.edu}
Chris~W.~Purcell$^{2}$,%\thanks{E-mail: cwpsps@rit.edu}
\newauthor
Frank~C.~van~den~Bosch$^{3}$, %\thanks{E-mail: frank.vandenbosch@yale.edu} 
Benedikt~Diemer$^{4}$, %\thanks{E-mail: benedikt.diemer@cfa.harvard.edu}
Johannes~U.~Lange$^{3,5}$,
Kuan~Wang$^{1}$, 
\newauthor
and Duncan Campbell$^{3}$
\vspace*{8pt}
\\
$^{1}$Department of Physics and Astronomy \& Pittsburgh Particle Physics, Astrophysics, and Cosmology centre (Pitt-PACC), \\
\phantom{$^{1}$}University of Pittsburgh, Pittsburgh, PA\\
$^{2}$School of Physics and Astronomy \& Centre for Computational Relativity and Gravitation (CCRG), Rochester Institute of Technology, Rochester, NY \\
$^{3}$Department of Astronomy, Yale University, PO. Box 208101, New Haven, CT 06520-8101\\
$^{4}$Institute for Theory and Computation, Harvard-Smithsonian centre for Astrophysics, 60 Garden St., Cambridge, MA 02138, USA \\
$^{5}$Kavli Institute for Theoretical Physics, University of California, Santa Barbara, CA}

\pubyear{2017}

\begin{document}
\label{firstpage}
\pagerange{\pageref{firstpage}--\pageref{lastpage}} 

\maketitle

\begin{abstract}
Dark matter halo clustering depends not only on halo mass, but also on other properties such as concentration and shape. This phenomenon is known broadly as {\em assembly bias}. We explore the dependence of assembly bias on halo definition, parametrized by spherical overdensity parameter, $\Delta$. We summarize the strength of concentration-, shape-, and spin-dependent halo clustering as a function of halo mass and halo definition. Concentration-dependent clustering depends strongly on mass at all $\Delta$. For conventional halo definitions ($\Delta \sim 200\mathrm{m}-600\mathrm{m}$), concentration-dependent clustering at low mass is driven by a population of haloes that is altered through interactions with neighbouring haloes. Concentration-dependent clustering can be greatly reduced through a mass-dependent halo definition with $\Delta \sim 20\mathrm{m}-40\mathrm{m}$ for haloes with $M_{200\mathrm{m}} \lesssim 10^{12}\, h^{-1}\mathrm{M}_{\odot}$. Smaller $\Delta$ implies larger radii and mitigates assembly bias at low mass by subsuming altered, so-called {\em backsplash} haloes into now larger host haloes. At higher masses ($M_{200\mathrm{m}} \gtrsim 10^{13}\, h^{-1}\mathrm{M}_{\odot}$) larger overdensities, $\Delta \gtrsim 600\mathrm{m}$, are necessary. Shape- and spin-dependent clustering are significant for all halo definitions that we explore and exhibit a relatively weaker mass dependence. Generally, both the strength and the sense of assembly bias depend on halo definition, varying significantly even among common definitions. We identify no halo definition that mitigates all manifestations of assembly bias. A halo definition that mitigates assembly bias based on one halo property (e.g., concentration) must be mass dependent. The halo definitions that best mitigate concentration-dependent halo clustering do not coincide with the expected {\em average} splashback radii at fixed halo mass.
\end{abstract}


%-----------------------
\section{Introduction}
\label{section:introduction}
%-----------------------

In the concordance cosmology, galaxies and clusters form within dark matter haloes \citep{white_rees78,blumenthal_etal84, mo_etal10}. Numerical simulations have provided a solid understanding of the abundances, properties, and clustering of dark matter haloes. Accordingly, it is possible to compute the clustering statistics of galaxies given a model for the relationship between galaxies and dark matter haloes. Such halo occupation models have been used to interpret large-scale structure measurements and constrain models of galaxy evolution \citep{yang_etal03,tinker_etal05,zehavi_etal05b,porciani_norberg06,vdbosch_etal07,zheng_etal07,conroy_wechsler09,yang_etal09b,zehavi_etal11,guo_etal11a,wake_etal11,yang_etal11a,yang_etal12,leauthaud_etal12,rodriguezpuebla_etal12, behroozi_etal13b, moster_etal13, tinker_etal13,cacciato_etal13,more_etal13,guo_etal14,zu_mandelbaum15b,desjacques_etal17}. To date, the vast majority of halo occupation models rely on a key assumption, namely that the probability a halo hosts a number of galaxies of a particular type depends only upon halo mass. It is now well known that the clustering strength of haloes depends upon other halo properties such as formation time \citep{gao_etal05,harker_etal06,wechsler_etal06,gao_white07,croton_etal07,zentner07,dalal_etal08, li_etal08, lacerna_padilla11}, concentration \citep{wechsler_etal06,faltenbacher_white10}, and other halo properties  \citep{bett_etal07, hahn_etal07a, hahn_etal07b, hahn_etal09, faltenbacher_white10, hester_tasitsiomi10, lacerna_padilla12, vandaalen_etal12, fisher_faltenbacher16, sunayama_etal16, chavesmontero_etal16}. If galaxy properties depend upon these halo properties, a phenomenon commonly referred to as galaxy {\em assembly bias}, then standard halo occupation modelling will fail \citep{zentner_etal14} and more complex models \citep{gilmarin_etal11, hearin_etal16} will be necessary for certain purposes.

There exist hints of observational detection of assembly bias in the literature. Using galaxy-group cross correlations, \citet{yang_etal06} have shown that at fixed mass haloes with quenched central galaxies are more strongly clustered than haloes hosting star forming centrals \citep[see also][]{blanton_berlind07, wang_etal08, wang_etal13}. Additional support for galaxy assembly bias comes from the detection of galactic conformity \citep{weinmann_etal06}, which indicates that the star formation properties of galaxies are correlated over scales significantly larger than the virial radii of their host haloes \citep[2-halo conformity][]{kauffmann_etal13,kawinwan_etal16,berti_etal17}. Although potentially indicative of galaxy assembly bias \citep[e.g.,][]{hearin_etal15,hearin_etal16}, some studies have argued that 2-halo conformity instead arises from contamination in the isolation criteria used \citep{tinker_etal17, sin_etal17}. \citet{miyatake_etal16} and \citet{more_etal16} presented  evidence that clusters with more centrally-concentrated satellite galaxy distributions cluster more weakly than  their less centrally-concentrated counterparts (however \citealt{zu_mandelbaum16} demonstrate that this claim is likely to be compromised by environment-dependent cluster membership assignment). 

Additional hints of galaxy assembly bias come from efforts to infer statistically the parameters of the galaxy--halo relationship from galaxy clustering data. In particular, \citet{lehmann_etal17} demonstrated that observed galaxy clustering is better described by models in which galaxy stellar mass depends upon halo concentration in addition to halo mass. Using decorated halo occupation distribution (HOD) models \citep{hearin_etal16}, a generalization of standard HOD models, \citet{zentner_etal16} interpreted observed galaxy clustering in the context of models that include assembly bias, finding marginal evidence that halo occupation depends upon both halo mass {\em and} concentration. These findings suggest that assembly bias is important to consider when interpreting galaxy clustering and that assembly bias may be critical to the interpretation of forthcoming, higher-precision data sets.

In this work, we explore the possibility of simplifying halo occupation modelling, at least for some applications, by altering the definition of the boundary of a halo. Though motivated broadly by spherical collapse \citep{gunn_gott72, fillmore_goldreich84, ryden_gunn87, lacey_cole93, eke_etal96, mota_vandebruck04, pace_etal10}, specific halo definitions have become a matter of convention that vary considerably in the literature. Many authors define haloes using a friends-of-friend (FoF) algorithm applied to the particle distribution \citep[e.g., ][]{davis_etal85}. More often, authors define haloes by spherical overdensity (SO) regions within which the mean density exceeds a particular threshold \citep[e.g., ][]{lacey_cole93}. These halo definitions are reviewed in \citet[][and references therein]{knebe_etal11}. The threshold used varies significantly. With respect to the mean density of the universe, commonly used thresholds are 178m, 180m, 200m, and $\sim$340m times the mean background density, motivated by spherical collapse models for different cosmologies. Many authors define haloes by an overdensity of 200c with respect to the critical density of the universe, motivated as an approximation to the virial radius at high redshift. This corresponds to an overdensity of 625m with respect to the mean density in a concordance cosmological model. Significantly higher values of the overdensity parameter are often used in studies of X-ray emission from cluster-sized haloes, where observations only cover large overdensities and correspondingly small radii.

We study the strength of various halo assembly bias signals as a function of halo definition. To be specific, we restrict ourselves to haloes that are defined by spherical regions and we study the strength of halo assembly bias as a function of the density threshold used to demarcate these haloes. This is motivated, in large part, by recent literature suggesting that a large portion of assembly bias stems from haloes in the relatively dense environments surrounding larger haloes \citep{wang_etal07,warnick_etal08,hahn_etal09,ludlow_etal09,lacerna_padilla11,more_etal15,sunayama_etal16}. The picture is that interactions with a relatively small number of large haloes influence the evolution and structural properties of nearby, smaller haloes. Indeed, the environmental impacts of haloes on one another has been shown to extend well beyond traditional virial radii \citep{wetzel_etal14, diemer_kravtsov14, behroozi_etal13b, adhikari_etal14, wetzel_nagai15, more_etal15, hearin_etal16}. Haloes in dense environments (e.g., near other large haloes) exhibit anomalous properties (e.g., formation times, concentrations, shapes) compared to field haloes in part because of their interactions with their larger, neighbours. These harassed haloes are referred to as {\em backsplash} haloes in much of the literature. It is interesting to ask whether or not a halo definition in which large haloes contain many of these smaller, anomalous, neighbour haloes can mitigate halo assembly bias. In this way, the halo boundary may be defined to serve as a more meaningful delineation of regions within which highly nonlinear effects are important. Assessing this strategy is an aim of our paper. 

We conclude that alternative halo definitions do {\em not} generally mitigate halo assembly bias for two reasons. First, no single halo definition mitigates all forms of halo assembly bias. For example, a halo definition that mitigates concentration-dependent halo clustering does not simultaneously mitigate angular momentum-dependent halo clustering. Second, any halo definition that mitigates any form  of assembly bias must be mass dependent. This suggests that assembly bias cannot be eliminated simultaneously at all halo masses without introducing a complex halo identification procedure. The details of our work have other interesting implications. Perhaps most importantly, our work emphasizes the fact that halo assembly bias effects are strongly dependent upon halo definition. Therefore, one must exercise care in comparing the work of different authors who may have used different halo definitions. Our final results summarizes halo assembly bias as a function of halo definition.
 
In Section~\ref{section:data} of this paper, we discuss the cosmological simulations and the halo finder utilized in the analysis. In Section~\ref{section:haloprops}, we discuss and define the halo properties used as tracers of assembly bias. In Section~\ref{section:methodology}, we discuss the statistics that we use to measure environmental effects after halo redefinition. We also discuss our method of removing known mass scaling from halo properties. In Section~\ref{section:results}, we present our findings and consider how the change of halo definition impacts measures of assembly bias. In Section~\ref{section:discussion}, we draw broad conclusions from this study of halo definition, with focus given toward understanding mechanisms that may drive our results. In Section~\ref{section:conclusions}, we discuss the significance of our results in the context of halo modelling. We also consider the nature of assembly bias as a function of halo definition.

%------------------------------------
\section[]{Simulations and Halo Identification}
\label{section:data}
%------------------------------------

In order to study the effects of halo redefinition, we use three cosmological $N$-body simulations of structure formation. These simulations are a subset of the  \citet{diemer_kravtsov15} simulations performed within the Planck best-fit cosmology with $\Om = 0.32$, $\Ol = 0.68$, $h_0 = 0.67$, $\sigma_8 = 0.834$, and $n_s = 0.9624$. We use three simulation boxes with comoving box lengths of 125, 250, and 500$\hMpc$ respectively. Each simulation models the evolution of $1024^3$ particles implying particle masses of $1.6 \times 10^8$, $1.3 \times 10^9$, and $1.0 \times 10^{10}\hMsun$ respectively. The three simulations have force softening scales of $2.4$, $5.8$, and $14 \hkpc$. We refer to each simulation as \simA, \simB, or \simC~ for the remainder of the paper. This set of simulations allows us to probe resolution effects and the mass dependence of halo clustering over a wider range of halo masses than would be possible with only one simulation. For example, \simA, with its higher resolution, contains the least massive resolved haloes, while \simC~has a larger number of high-mass haloes.

%------------- mass as a function of Delta
\begin{figure}
\centering
\includegraphics[width=\columnwidth]{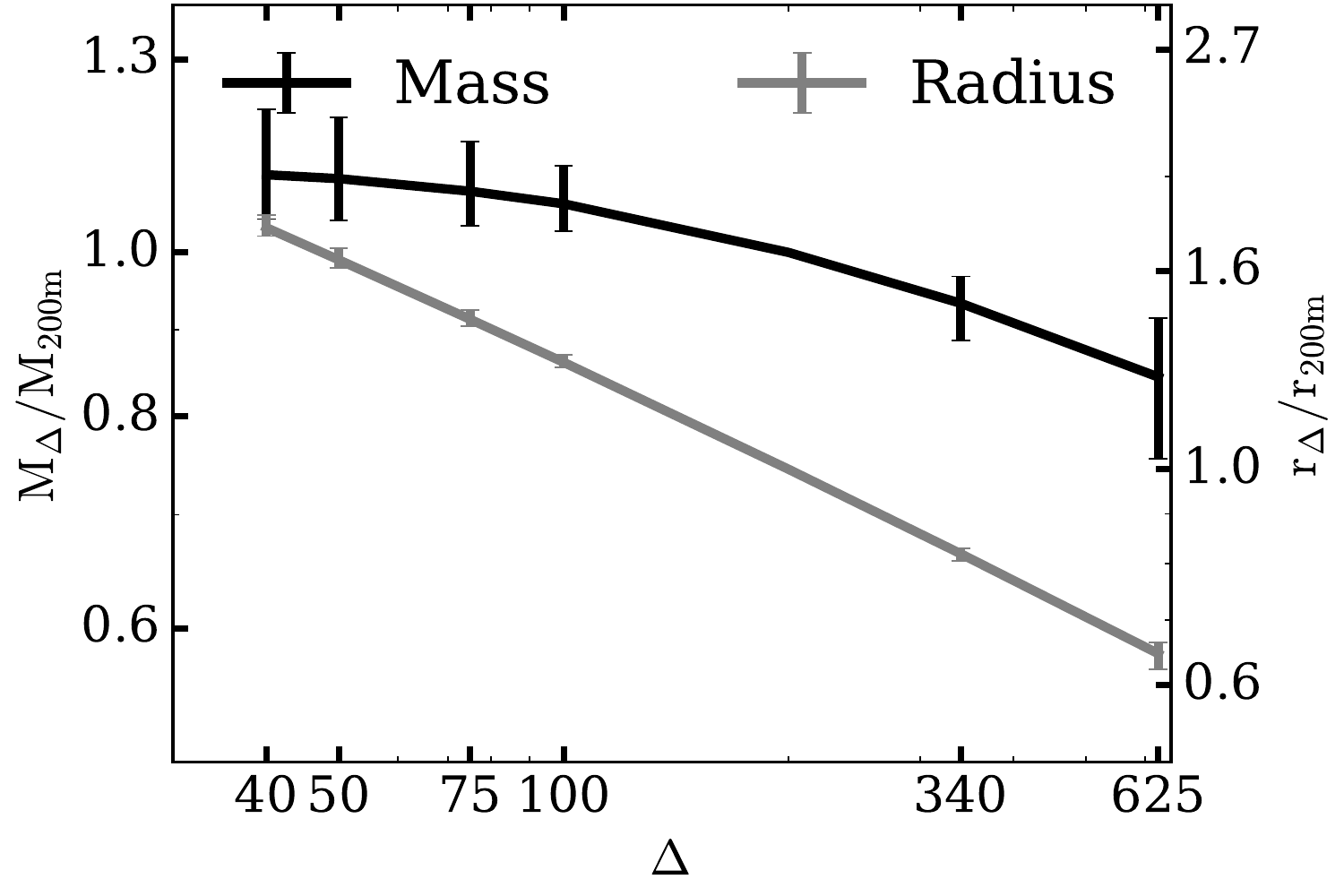}
\caption{
The ratio of halo properties as a function of $\Delta$ in the \simB~ catalogue. The sample contains all host haloes with masses greater than $M_{200\mathrm{m}} \ge 7.1 \times 10^{11}$. The black (dark grey) line shows the median value of the ratio of the halo mass (halo radius) at a value of $\Delta$ to the value at $\Delta=200\mathrm{m}$. The error bars contain 68\% of values of this ratio for the sample. These ratios are 
very mild functions of mass, but all masses are stacked in this plot.}
\label{fig:deltacompare}
\end{figure}
%-------------------------------------

To identify haloes, we use the {\tt ROCKSTAR} halo finder, which works on the phase space algorithm described in \citet{behroozi_etal13a}. In short, {\tt ROCKSTAR} determines initial groupings of particles using a Friends-of-Friends (FOF) algorithm in phase space before applying the spherical overdensity halo definition in order to calculate halo properties of interest. Unbound particles are removed prior to the calculation of halo mass and other halo properties. A halo is given a radius, $\Rdel$, determined by
\beq
\label{eq:delta-def}
	\bar{\rho}(\Rdel) = \Delta \rho_{\mathrm{m}}, 
\eeq
where the mean density within a spherical volume of radius $r$ is $\bar{\rho}(r)$, $\Delta$ is the overdensity parameter, and $\rho_{\mathrm{m}}$ is the mean background mass density of the simulation. 
Some authors use critical density  $\rho_{\rm crit}$ in Eq.~\eqref{eq:delta-def} instead of $\rho_{\mathrm{m}}$. Because $\rho_{\mathrm{m}} = \Omega_{\rm m}\rho_{\rm crit}$, this difference leads to significant difference in halo definition if the critical density is used. Following a common convention, we add the suffix ``m'' or ``c'' after the value of the overdensity $\Delta$, to indicate whether the overdensity is with respect to the mean density (as in our work) or the critical density, respectively.

The {\tt ROCKSTAR} software also effectively identifies halo substructure, commonly referred to in the literature as subhaloes. All density peaks are identified within the simulation and if a halo center exists within the FoF group of another halo, the less massive of the two is defined to be a subhalo of the more massive companion. The more massive companion is referred to as the host halo. This process continues until all haloes identified in the simulation have been designated as either host haloes or subhaloes.

The {\tt ROCKSTAR} software also effectively identifies halo substructure, commonly referred to in the literature as subhaloes. All density peaks are identified within the simulation and are assigned as seed halos. Particles are then assigned using a phase-space algorithm that minimizes misassignment on the outskirts of the halos. Satellite membership of the largest seed halos is determined using these phase-space distances as well in order to minimize ambiguity in substructure assignment. After this internal step, host/subhalo membership is reassigned according to a standard definition (e.g., subhaloes are within $\Rdel$ of more massive host halos). This method allows us to know that subhalo identification remains robust even as we push the halo finder to extreme values of $\Delta$.

Halo radii vary widely dependent upon the choice of $\Delta$%
\footnote{It is worth noting that the {\tt ROCKSTAR} linking length parameter must be adjusted as $\Delta$ varies in order to ensure that SO haloes contain all relevant particles. The value we choose ($0.4$) is large enough that it works for all definitions, though unoptimized. Our analysis shows no spurious behaviour due to this choice.}. The number chosen for $\Delta$ varies throughout the literature, usually from $\Delta \approx 178$m to $\Delta \approx 340$m. Some authors define haloes as an overdensity of 200m while others use 200c ($\approx$ 625m). Many use the ``virial'' overdensity \citep{Bryan1998}, which in a flat $\Lambda$CDM cosmology with $\Omega_M \simeq 0.3$ is about 100c (or $\approx$ 300m). X-ray studies of clusters use higher overdensities to isolate the regions within which they have data, with overdensities reaching as high as $\Delta \approx 7800$m ($\approx 2500$c).

We vary the sizes of a haloes by treating the overdensity parameter as tunable, and our most interesting results correspond to values from $\Delta \approx 20$m to $\Delta \approx 625$m, which spans the range used in the literature on galaxy clustering. The impact of the choice of $\Delta$ on halo radius and halo size can be seen in Fig.~ \ref{fig:deltacompare}. As one should expect, as $\Delta$ increases, the halo boundary corresponds to a smaller sphere containing less mass. 

%------------------- Cartoon Figure that shows halo redefinition
\begin{figure*}
\centering
\includegraphics[width=0.7\textwidth]{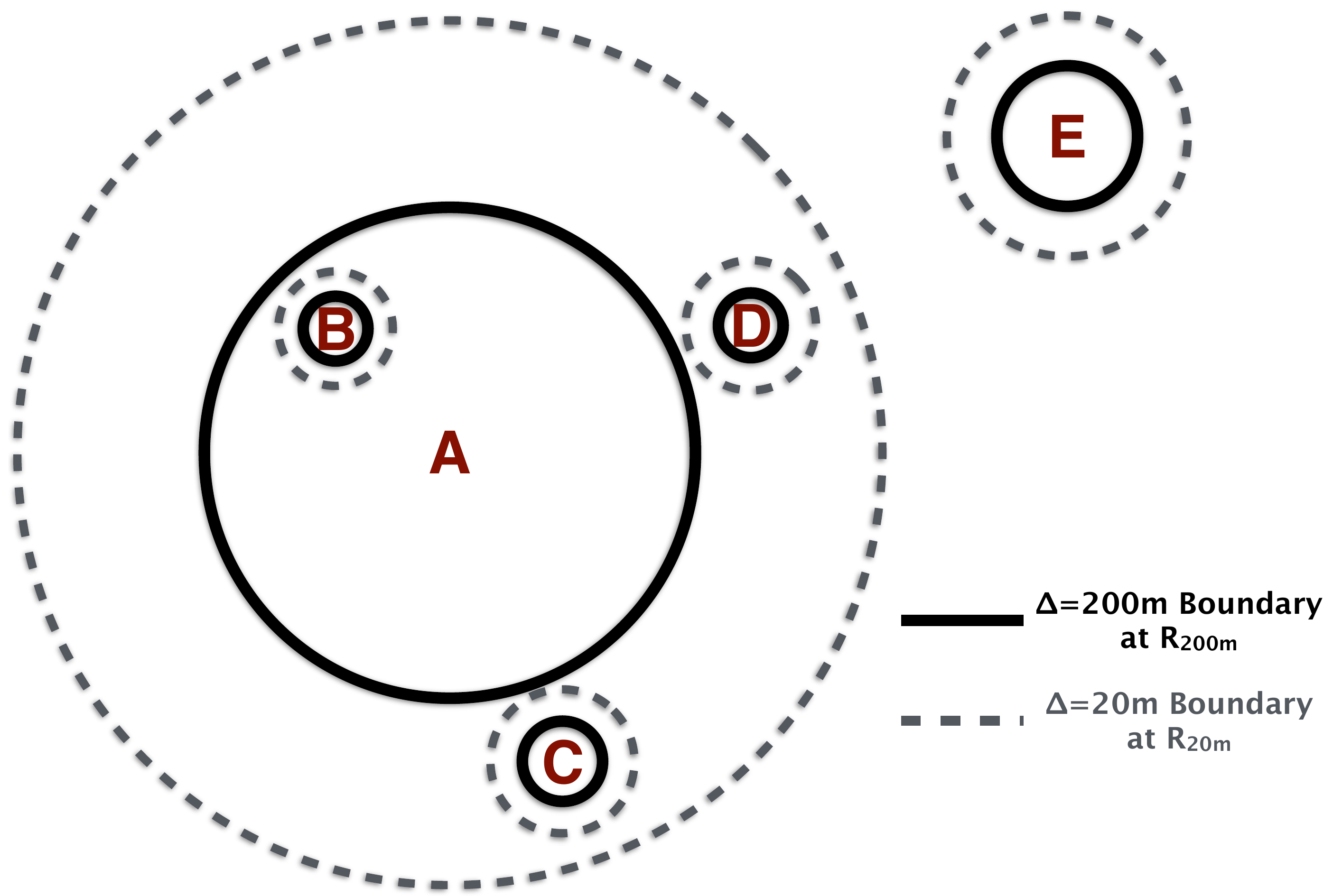}
\caption{
A qualitative illustration of halo redefinition. The figure shows five haloes, labelled by the letters A-E. Halo A is the largest halo in the illustration. The solid halo boundaries correspond to the halo radii defined with respect to an overdensity of $\Delta=200$m, namely $R_{200\text{m}}$. The dashed boundaries correspond to the halo radii defined with respect to an overdensity of $\Delta=20$m, $R_{20\text{m}}$. In all cases, haloes become larger as $\Delta$ decreases. Haloes A and E are host haloes according to both halo definitions. Halo B is a subhalo of halo A according to both halo definitions. Haloes C and D are distinct host haloes according to the $\Delta=200$m halo definition, but they are reclassified as subhaloes of halo A according to the $\Delta=20$m halo definition.
}
\label{fig:halocartoon}
\end{figure*}
%----------------------------------

To build an intuitive picture of the manner in which halo definitions may alter halo properties and halo clustering, consider the cartoon in Figure~\ref{fig:halocartoon}. This figure shows five haloes, labelled with the letters A through E. Halo A is the largest of these haloes. The figure shows the haloes, with boundaries that are defined with respect to two overdensity criteria, $\Delta=200$m and $\Delta=20$m. Decreasing $\Delta$ results in increasing halo radii and masses, which is illustrated by the difference between the solid boundaries in Fig.~\ref{fig:halocartoon} (which correspond to $R_{200{\rm m}}$) and the dashed boundaries (which correspond to $R_{20{\rm m}}$). This is quantified in Fig.~\ref{fig:deltacompare}. For any individual halo, the change in mass as $\Delta$ is decreased depends upon the environment of the halo. In Fig.~\ref{fig:halocartoon}, the material between the solid and dashed radii is material that is in the environment of the $\Delta=200$m haloes but is subsumed within the $\Delta=20$m haloes. This leads to the scatter shown in Fig.~\ref{fig:deltacompare}. 

Another important feature illustrated by Fig.~\ref{fig:halocartoon} is that the classification of a halo as either a host halo or a subhalo depends upon halo definition. Haloes A and E are host haloes according to both definitions, while halo B is a subhalo in
both cases. Haloes C and D, on the other hand, are considered distinct host haloes when haloes are defined according to an overdensity of $\Delta=200$m, but subhaloes when $\Delta=20$m. If haloes C and D have properties that are strongly altered by interactions with halo A, while haloes A and E do not have properties that have been significantly altered by their mutual interaction, then the $\Delta=20$m halo definition may mitigate assembly bias and may serve as a more practical halo definition. This is part of what we explore in this paper.

It is worth emphasizing that the clustering of host haloes can vary for multiple reasons as a function of halo definition. First, different halo definitions lead to different halo properties because of the mass included or excluded by the particular definition. Second, the sample of host haloes itself varies with halo definition. According to the $\Delta=200$m definition, Fig.~\ref{fig:halocartoon} shows four distinct host haloes, namely A, C, D, and E. However, according to the $\Delta=20$m definition, Fig.~\ref{fig:halocartoon} shows only two host haloes, A and E. Decreasing $\Delta$ results in fewer distinct host haloes. We discuss these points in greater detail in Section~\ref{section:discussion}.

When it is required to compare individual host haloes across different values of $\Delta$, we compare all halo centers in the source catalogue to the halo center of each host halo in the target catalog. Here both the source and target catalogs are output by {\tt ROCKSTAR}, but only with different halo definitions. We find the nearest neighbour to each host halo in the target catalogue and determine if it is within a small tolerance, set to be 1\% of the halo radius as defined by the source catalog. While this ultimately will result in some degree of mismatching, we believe that these should not have strong impact on the resulting summary statistics.

%-------------------------------------
\section{Halo Properties}
\label{section:haloprops}
%-------------------------------------

As has long been well known, halo mass is the property that most strongly affects halo clustering. In this paper, we aim to study the strength of halo clustering as a function of halo properties other than mass. We will refer to these properties as ``auxiliary'' halo properties in this paper. Throughout the remainder of the paper, we refer to halo two-point clustering that depends upon one of these auxiliary properties at fixed mass variously as ``halo assembly bias,'' or ``auxiliary property dependent" clustering. The properties we study are described in this section.

%----------------------------------------
\subsection{Measures of Halo Concentration}

We investigate halo clustering as a function of two definitions of halo concentration. The first stems from a fit of the spherically-averaged halo density profile, $\rho(r)$, to a \citet[hereafter NFW]{navarro_etal97} density profile, 
\beq
\rho(r) = \frac{\rho_0}{\frac{r}{r_{\mathrm{s}}}\left(1+\frac{r}{r_{\mathrm{s}}}\right)^2},
\eeq
where the density scale, $\rho_0$, and the scale radius, $r_{\mathrm{s}}$, are parameters that are fit to the density profile of each halo. The standard definition of halo concentration is then 
\beq
c_{\mathrm{NFW}} = \frac{\Rdel}{r_\mathrm{s}},
\eeq
where $\Rdel$ is the radius of the halo given an overdensity parameter, $\Delta$, defining the halo and $r_{\mathrm{s}}$ is the inferred halo scale radius. We use the concentration parameter in the {\tt ROCKSTAR} halo catalogues, which is obtained by fitting the scale radius of the NFW profile to the radial density profile of each halo, whose centre is the average positions of about the 0.1\% innermost particles \citep{behroozi_etal13a}. 

The NFW concentration defined in the previous paragraph has the shortcoming that it depends upon a parametric description of dark matter haloes. As an alternative, we therefore also study the clustering dependence of haloes as a function of a non-parametric description of halo concentration. In particular, we use the ``velocity-ratio'' concentration \citep{prada_etal12,klypin_etal16},
\beq
c_{\mathrm{V}} = \frac{V_{\mathrm{max}}}{V_{\Delta}}, 
\eeq
where $V_{\mathrm{max}}$ is the maximum circular velocity achieved within the halo and $V_{\Delta}$ is the circular velocity at the halo radius, $\Rdel$. All haloes of the same mass have the same value of $V_{\Delta}$; however, they exhibit a variety of $V_{\mathrm{max}}$ values depending upon the degree to which their masses are concentrated toward the halo centre. The quantity $c_{\mathrm{V}}$ is a non-parametric measure of halo concentration and can be measured from simulation snapshots without fitting halo density profiles. Consequently, $c_{\mathrm{V}}$ is robust to halo density profile parametrizations and halo profile fitting procedures. If an NFW profile is assumed, one expects the relationship between $c_{\rm V}$ and $c_{\rm NFW}$ to follow a simple relation,
\beq
c_{\rm V} = 0.465 \left[\frac{\ln\left(1+c_{\rm NFW}\right)}{c_{\rm NFW}}-\frac{1}{1+c_{\rm NFW}}\right]^{-1/2}.
\eeq
In light of halo assembly bias, halo concentrations are interesting for a number of reasons. First of all, the environment dependence of halo concentrations is known to be strong for standard halo definitions. Second, halo concentrations are of interest in the modelling of galaxy clustering and gravitational lensing statistics (and their cross correlations). In the case of galaxy clustering, the relevance is indirect, because satellite galaxies in groups may not trace the mass densities of their host haloes. In the case of gravitational lensing, the mass distribution is the primary quantity of interest and halo concentrations are directly related to predictions for lensing statistics.

A third motivation to study halo concentrations is that halo concentrations are known to be strongly correlated with the formation histories of dark matter haloes with earlier forming haloes having higher concentrations at fixed halo mass \citep{wechsler_etal02, zhao_etal03, wechsler_etal06, zhao_etal09}. As such, exploring the concentration dependence of halo clustering may yield insight into the age dependence of halo clustering without the need for constructing merger trees. This is particularly important for our study in which the halo finding is performed repeatedly for many different values of $\Delta$. Constructing a self-consistent merger tree from which halo formation history can be studied requires halo finding at all simulation snapshots for each new value of $\Delta$, which is a computationally expensive task for the purposes of the present exploratory study. In the present paper, we limit our study to halo properties that can be measured from a single simulation snapshot and study haloes at $z=0$.

A note of caution regarding the preceding discussion is that halo formation histories correlate with halo concentrations with significant scatter and this correlation may depend upon environment, so the reader must be wary of drawing conclusions about the environmental dependence of halo formation by extrapolating our results on halo concentration. We will explore measures of halo age directly in a forthcoming follow-up study dedicated to halo formation histories.

One final motivation is that halo concentration can be sensitive to halo definition in a non-trivial manner that may influence concentration-dependent clustering. As demonstrated in \citet{wechsler_etal02}, the clustering dependence of halo concentration exhibits a mass dependence which includes a sign change near the collapse mass. Further, as our cartoon in Fig.~\ref{fig:halocartoon} demonstrates, subsumed (sub)halo particles can alter halo density profiles and inferred halo concentrations. This is particularly true because subhaloes preferentially reside in the outer regions of halos. It is possible that this renders concentration more sensitive to halo definition than other halo properties and we will return to this question in Section~\ref{section:discussion}.

\begin{figure*}
\centering
\includegraphics[width=\textwidth]{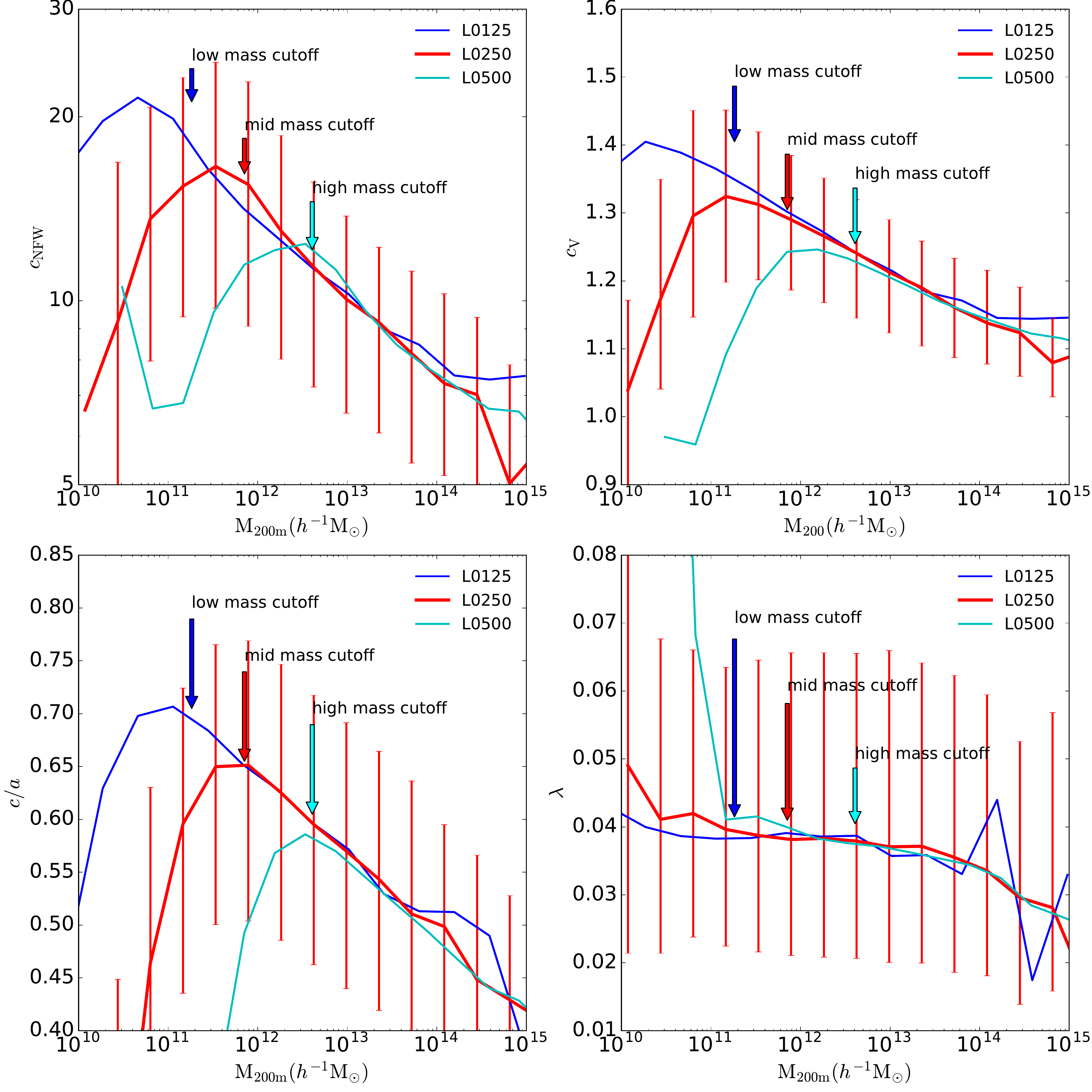}
\caption{
The relationship between mean halo properties and halo mass for each of our simulations with $\Delta =200$m. These properties are the halo NFW-defined concentration, the velocity ratio defined concentration, the halo shape, and the halo spin. For example, in the top left panel, in order of increasing simulation volume, the blue line corresponds to the concentration-mass relation from simulation \simA, the red line corresponds to \simB, and the cyan line corresponds to \simC. The red error bars show the 68\% spread in parameter values within that mass bin for \simB. These errors are comparable to those of the other simulations within the region of interest. Each simulation is subject to resolution limitations at different halo masses. We show with arrows the minimum $M_{200\text{m}}$ mass thresholds that we adopt in our analyses using the same colour code as the concentration-mass relations, going from \simA \ to \simC \ from left to right. Note the deviation from monotonic trends as a result of resolution effects.
}
\label{fig:massrelation}
\end{figure*}

\begin{figure}
\centering
\includegraphics[width=\columnwidth]{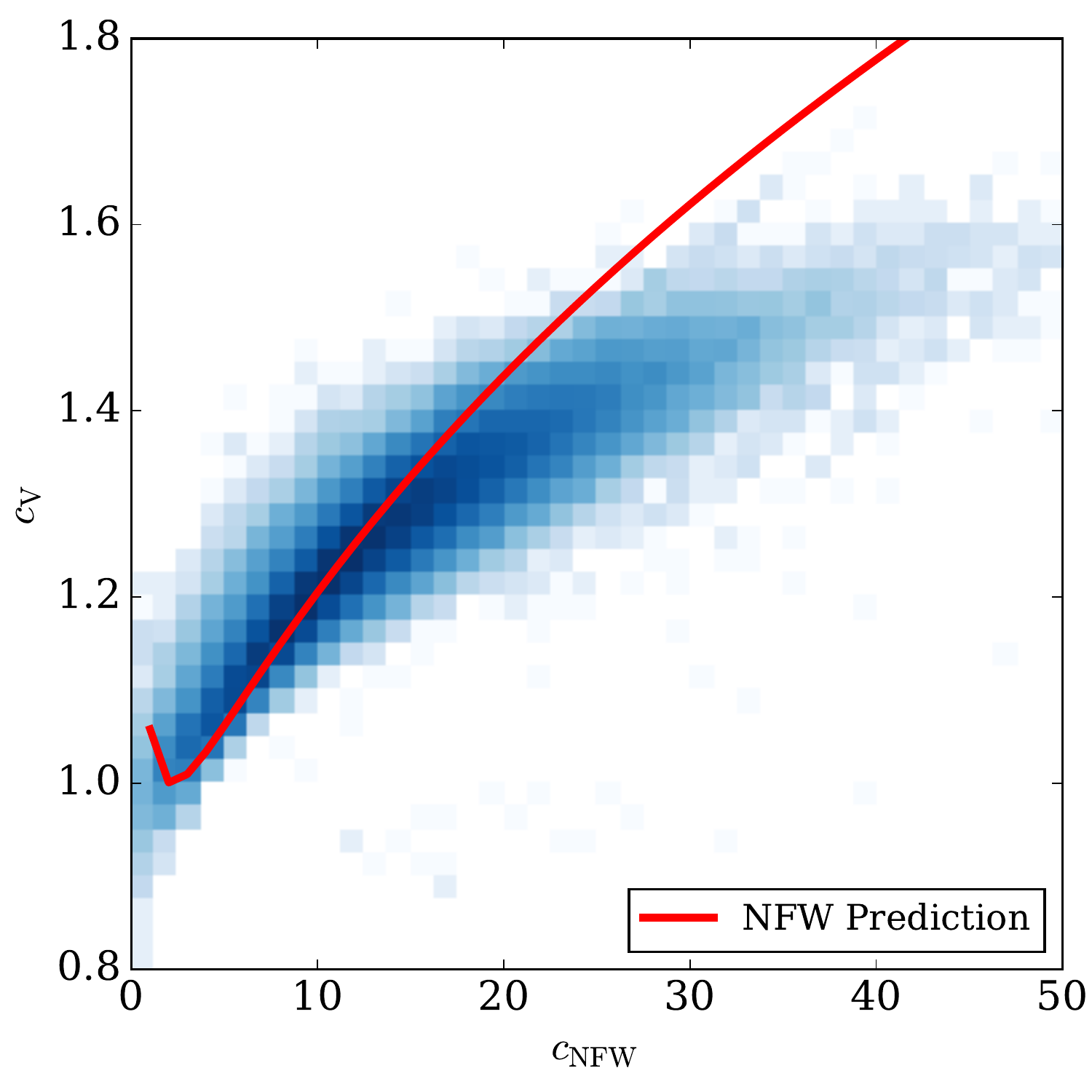}
\caption{
The relationship between the two different measures of concentration, using haloes in \simB~ defined with $\Delta=200$m. The colour scale, shown at the right, encodes the number of haloes within a single two-dimensional bin in the $c_{\mathrm{NFW}}-c_{\mathrm{V}}$ space. The dark (light) blue regions on the plot show where the most (fewest) haloes exist with those values of the two concentration parameters. The white regions indicate where no haloes hold these values. The scatter on this relationship ranges from 5\% for intermediate concentration values, to a high of 13\% at high masses. The red line shows the prediction from  assuming an NFW profile.}
\label{fig:concentrations}
\end{figure}

The upper-left panel of Fig.~\ref{fig:massrelation} shows the mean $c_{\mathrm{NFW}}$-$M_{\Delta}$ relation for haloes defined with $\Delta=200$m in \simA, \simB, and \simC. For each simulation, we consider haloes only above a minimum mass threshold to ensure that property measurements are not compromised by resolution effects. The minimum mass thresholds are shown as the downward-directed arrows in Fig.~\ref{fig:massrelation} and are listed in Table~\ref{table:thresholds} alongside the associated minimum number of particles\footnote{While these choices of minimum particle number are not terribly conservative compared to those from the literature, we note that our choice of normalization below mitigates most of the impact of those poorly resolved halos in the sample.}. The upper-right panel of Fig.~\ref{fig:massrelation} also shows the relation between the velocity-ratio concentration, $c_{\mathrm{V}}$, and halo mass. Halo concentration is a decreasing function of halo mass, a result that is consistent with the significant previous literature on the subject \citep[e.g.,][and references therein]{bullock_etal01, maccio_etal07,duffy_etal08,prada_etal12,klypin_etal16}.  Fig.~\ref{fig:concentrations} shows the relationship between $c_{\mathrm{NFW}}$ and $c_{\mathrm{V}}$ on a halo-by-halo basis. As is evident, the two concentration proxies are strongly correlated and exhibit a $\sim 6\%$ scatter indicating that $c_{\mathrm{NFW}}$ and $c_{\mathrm{V}}$ indeed encode similar information about each halo.

%------------------------------------------ Fig.~ for Cnfw(M)
%--------------------------------------------------------------------------------

%-------------------------------------------- Fig.~ comparing Cnfw and Cv on a halo-by-halo basis
%--------------------------------------------------------------------------------------------------------------------------------

%---------------------------------------
\subsection{Halo Shape}

In addition to halo concentrations, we examine halo clustering as a function of a variety of other halo properties. We study halo clustering as a function of halo shape, $s$, quantified by the ratio of the halo minor, $c$ and major, $a$, axis lengths, 
\beq
s = \frac{c}{a}.
\eeq
The halo shapes we used were measured by {\tt ROCKSTAR} using the method in \cite{allgood_etal06}, which calculates a modified inertia tensor,
\beq
M_{ij}=\frac{1}{N}\sum\limits_{k=1}^{N}\frac{x_{i,k} x_{j,k}}{r_{k}^2}\,.
\eeq
Here $x_{i,k}$ is the location of particle $k$ along axis $i$ with respect to the halo centre and $r_{k}$ is the distance between the particle and the halo centre. The factor $1/r_{k}^2$, which is not part of the standard definition of the inertia tensor, is included in order to diminish the influence of massive subhaloes with large halocentric distances. on the measurement of the halo shape. The calculation starts with using all particles within a sphere of the halo radius; the shape of boundary then changes iteratively according to the shape measurement of each iteration. Particles associated with identified substructure are included in the calculation. The sorted eigenvalues of the matrix represent the squares of the principal ellipsoid axes, where $a > b > c$. 

The mean relations for halo shapes as a function of halo mass for $\Delta=200$m are shown in the lower left panel of Fig.~\ref{fig:massrelation} along with the mass thresholds used to ensure that our results are not compromised by resolution. The relationship between halo shape and halo mass is such that haloes become somewhat less spherical as halo mass increases in accord with previous studies of halo shapes \citep[e.g.,][]{jing_suto02,allgood_etal06}.

%-----------------------------
\subsection{Halo Spin}

We study halo clustering as a function of halo angular momentum quantified by the spin parameter, $\lambda$, as introduced by \citep{peebles69},
\beq
\lambda = \frac{J \sqrt{\lvert E\rvert}}{G M_{\Delta}^{2.5}}
\eeq
where $J$ is the halo angular momentum, $E$ is the total energy of the halo, and $M_{\Delta}$ is the mass enclosed by the halo radius, $r_{\Delta}$. Spin is a measure of the angular momentum of the halo in units of the angular momentum necessary to support the halo against collapse entirely by rotation. We measure this quantity using {\tt ROCKSTAR} which calculates the angular momentum, total energy, and total mass within $\Delta$ using bound particles out to the corresponding halo radius. The mean relations for halo spin as a function of halo mass for $\Delta=200$m are shown in the lower right panel of Fig.~\ref{fig:massrelation} along with the mass thresholds (shown in Table~\ref{table:thresholds}) enforced to ensure that our results are not compromised by resolution. Halo spins are typically $\lambda \sim 0.04$ and exhibit only a very weak mass dependence in accord with independent studies of halo spin 
\citep[e.g.][]{bullock_etal02,maccio_etal07}.

As with concentration-dependent clustering, we may expect spin-dependent 
halo clustering to depend upon halo definition as subhaloes are subsumed 
into larger host halos as overdensity threshold decreases. This potential sensitivity to halo definition partly motivates the study of halo spin-dependent clustering.

%---------------------------------
\subsection{Halo Samples}

In practice, the mean relations between the various halo properties and the mass 
thresholds for our analyses must be determined separately for each combination of simulation, halo property (e.g., $c_{\mathrm{NFW}}$, $c_{\mathrm{V}}$, $\lambda$, or $s$), and halo definition (i.e., value of $\Delta$). For each analysis, we set mass thresholds in order to avoid the regime in which halo parameters are not well measured due to resolution limits of the simulations; we draw attention to the deviation in the upper-left panel of Fig.~\ref{fig:massrelation} as an example of this. The NFW defined halo concentration follows an approximate power law with halo mass; however, at low particle numbers it is difficult to fit to an NFW profile and we see significant deviation from the mean relation. Simulations with larger number of particles at lower masses do not demonstrate this same effect, implying that this is primarily a function of simulation resolution.

For ease of comparison between halo definitions, we choose to use a single mass threshold for each simulation and for each value of $\Delta$. The mass thresholds are chosen to simultaneously minimize resolution effects and to include a similar population of halos%
\footnote{In particular, the mass thresholds have been chosen to minimize the amount of halos that are removed or added when moving from one halo definition to the next solely due to a shift in mass cutoffs.} above the threshold for each value of $\Delta$. We summarize the mass thresholds we have used for a subset of $\Delta$ values in Table~\ref{table:thresholds}. Note that because halo abundance is a rapidly-declining function of halo mass, the statistics we quote are always dominated by halos near the lower edge of the mass range. We have recomputed all statistics with mass threshold samples (rather than mass bins) and find results that are both qualitatively and quantitatively similar. At most values of $\Delta$, the minimum mass thresholds are driven by the requirement that the halo properties do not suffer significantly from finite resolution effects. We alert the reader to the fact that the mass of an individual halo will vary as $\Delta$ varies. This effect can be seen in Fig.~\ref{fig:deltacompare}, which demonstrates that while a decreased value of $\Delta$ leads to larger masses on average, there is a scatter due to changes in halo identification. Roughly speaking, the threshold masses in Table~\ref{table:thresholds} vary in such a way that the 
same physical objects are selected at each halo definition.

%%%%%%%%%%%%%%%%%%%%%%%%%%%%%%%%%%%%%%%%%%%%%%%%%%%%%%%%%%%%%%%%%%%%%%%%%%%%
\begin{table*}
\caption{
Minimum mass thresholds for each of our analyses. 
In the columns below each value of $\Delta$, we show the minimum 
host halo masses considered in units of $h^{-1}\mathrm{M}_{\odot}$, the associated minimum number of particles, and the total number of host halos identified between this lower mass threshold and an upper mass threshold to avoid overlapping mass bins.
}
\vspace*{8pt}
\begin{tabular}{c | r | c c c c c c c }
\hline \hline
Simulation & & $\Delta=625$m & $\Delta=340$m & $\Delta=200$m & $\Delta=100$m & $\Delta=75$m & $\Delta=50$m & $\Delta=20$m \\ 
\hline
{\simA} & mass cut & $1.34 \times 10^{11}$ & $1.67 \times 10^{11}$ & $1.83 \times 10^{11}$ & $1.94 \times 10^{11}$ & $1.97 \times 10^{11}$ & $2 \times 10^{11}$ & $2.03 \times 10^{11}$  \\
 & \# particles & 837 & 1043 & 1143 & 1212 & 1231 & 1250 & 1268 \\
 & \# halos & 33623 & 29572 & 28467 & 27938 & 27754 & 27454 & 25329 \\
\hline
{\simB} & mass cut & $5.23 \times 10^{11}$ & $6.49 \times 10^{11}$ & $7.1 \times 10^{11}$ & $7.55 \times 10^{11}$ & $7.66 \times 10^{11}$ & $7.77 \times 10^{11}$ & {N/A} \\ 
 & \# particles & 402 & 499 & 546 & 580 & 589 & 597 &  \\
 & \# halos & 88045 & 79795 & 78465 & 78851 & 79344 & 79152 &  \\
\hline
{\simC} & mass cut & $2.99 \times 10^{12}$ & $3.71 \times 10^{12}$ & $4.06 \times 10^{12}$ & $4.31 \times 10^{12}$ & $4.38 \times 10^{12}$ & $4.44 \times 10^{12}$ & {N/A} \\ 
 & \# particles & 299 & 371 & 406 & 431 & 438 & 444 & \\
 & \# halos & 168742 & 158512 & 161339 & 169361 & 172175 & 175791 &  \\
\hline \hline
\end{tabular}
\label{table:thresholds}
\end{table*}
%%%%%%%%%%%%%%%%%%%%%%%%%%%%%%%%%%%%%%%%%%%%%%%%%%%%%%%%%%%%%%%%%%%%%%

%-----------------------
\section[]{Halo Clustering as a function of Auxiliary Halo Properties}
\label{section:methodology}
%-----------------------

%----------------------------------------------------------
\subsection{Auxiliary Halo Properties and Their Mass Dependences}
\label{subsection:properties}

We are interested in studying the clustering behaviour of haloes as a function of properties other than mass. We refer to those properties other than mass that we study as ``auxiliary halo properties." As has been demonstrated extensively in the literature, the auxiliary properties that we consider are themselves functions of mass.
This mass dependence, if not accounted for, induces clustering that depends upon these auxiliary properties even in the absence of assembly bias. Most contemporary cosmological $N$-body simulations, and specifically the suite of simulations that we study in this work, do not have a sufficiently large number of haloes to make isolating haloes of fixed mass, and then further splitting these haloes by an auxiliary property, a statistically powerful method with which to study the dependence of clustering on auxiliary properties. Therefore, it is necessary to remove and/or mitigate the mass dependence of the auxiliary properties that we study. 

We mitigate the mass dependence of the auxiliary properties by assigning haloes their auxiliary property marks as follows. First, we take all host haloes more massive than our minimum mass thresholds and sort them by their halo masses, $M_{\Delta}$. We place these haloes into 20 logarithmically-spaced bins of halo mass, ensuring that no bin has fewer than 10 haloes. We then calculate the rank of each auxiliary property within each bin of halo mass, from 1 to $N$, where $N$ is the number of haloes assigned to the bin. If multiple haloes share auxiliary property values (e.g., rank 2 and 3 have the same halo shape), then the average value of the rank is assigned to each halo (e.g., 2.5 to each). We normalize the rank distribution to be between 0 and 1 by dividing the ranks by the number of haloes in the bin, $N$. We have experimented with various bin sizes and binning schemes (e.g., equally populated bins, rather than evenly spaced bins) and these choices make no qualitative and little quantitative difference to any of our results.

In this manner, for each halo we replace the raw auxiliary properties with a percentile ranking for each property at fixed mass. For example, a halo assigned a $c_{\rm NFW}$ auxiliary property mark of $0.78$ has a concentration that is higher than $78\%$ of haloes at that mass. In this manner, the mass dependence of the marks is eliminated. An additional benefit is that the distribution of the auxiliary property marks is always a uniform distribution on the interval between zero and one. This proves convenient for the interpretation of marked correlation functions which we discuss below.

%---------------------------------------------------------------
\subsection{Clustering Statistics}
\label{subsection:clusteringstatistics}
%---------------------------------------------------------------

We assess the influence of assembly bias specifically on two-point statistics of host haloes. In order to do so, we study both the standard two-point correlation functions (CFs) of haloes selected by properties other than mass (e.g., the auxiliary properties concentration, shape, and spin) as well as halo mark correlation functions (MCFs). MCFs quantify the manner in which a halo property (the ``mark'') correlates among halo pairs as a function of the distance between the pairs. MCFs have the advantage that they effectively stack signal from all values of the halo auxiliary property, or mark, in contrast to selecting subsets of haloes based on the auxiliary property. MCFs also stack signal from all environments and do not require any specific definition of halo environment in order to detect ``environmental'' trends that are usually referred to as assembly bias in the literature. Absent halo assembly bias, the halo marks are uncorrelated among pairs. MCFs have been used in many previous papers to quantify environmental dependence of halo properties other than mass \citep{beisbart_kerscher2000,faltenbacher_etal02,sheth_tormen04,sheth05, skibba_etal06, harker_etal06,wechsler_etal06,mao_etal15}. Although it does not necessarily have to be the case, we find that using correlation functions of halo sub-samples and using MCFs lead to the same broad conclusions. 

For a specific halo property, or mark $m$, we use the MCF normalization of \citet{wechsler_etal06}, namely 
\begin{equation}
\label{eq:mcf}
\mathcal{M}_m(r) \equiv \frac{\langle m_i m_j \rangle_{\,i,j \in P(r)} - 
\langle m \rangle^2}{\operatorname{Var}(m)},
\end{equation}
where $\langle m_i m_j \rangle_{\,i,j \in P(r)}$ is the mean of the product of two marks of a pair of haloes separated by a distance about $r$, $\langle m \rangle$ is the mean of the mark for all haloes, and $\operatorname{Var}(m)$ the variance of the mark for all haloes.
In the absence of any correlation between a halo property among neighbours of a separation $r$ away, $\mathcal{M}_m(r) \simeq 0$ and its absolute value would be much less than 1. Deviations of the MCF from zero indicate such correlations exist and the magnitude of $\mathcal{M}_m(r)$ gives the excess of the mark among pairs compared to the one-point mean of the mark $\langle m\rangle$ in units of the one-point variance. The marks that we use are the normalized ranks of halo auxiliary properties described in Section~\ref{subsection:properties}, which are uniformly distributed between $\frac{1}{{\rm N}}$ and 1.

For each observable, it is necessary to assess the statistical significance of any auxiliary property dependent clustering signal. We assess the statistical significance of CFs and MCFs by random re-assignment of marks. For each halo we assign a randomized mark drawn from a uniform distribution between 0 and 1. Because no information about clustering is used in the assignment, the CFs and MCFs computed from these randomized marks can only exhibit auxiliary property dependent clustering to the degree allowed by finite sampling. We repeat this process 200 times and calculate the $2^{\mathrm{nd}}$ and $98^{\mathrm{th}}$ percentile values to determine the approximate 2 $\sigma$ intervals. Any measurement within this range is consistent with zero assembly bias and conversely, any measurement outside of this range is unlikely to arise due to finite sampling of the halo population and may be attributable to underlying halo assembly bias at $> 2\sigma$ significance.

%----------------------------
\section[]{Results}
\label{section:results}
%----------------------------

%-----------------------------------------------
\subsection{Correlation Functions}
\label{sub:cfresults}

%---------------------------------------------------------------------------------
\begin{figure*}
	\centering
	\includegraphics[width=\textwidth]{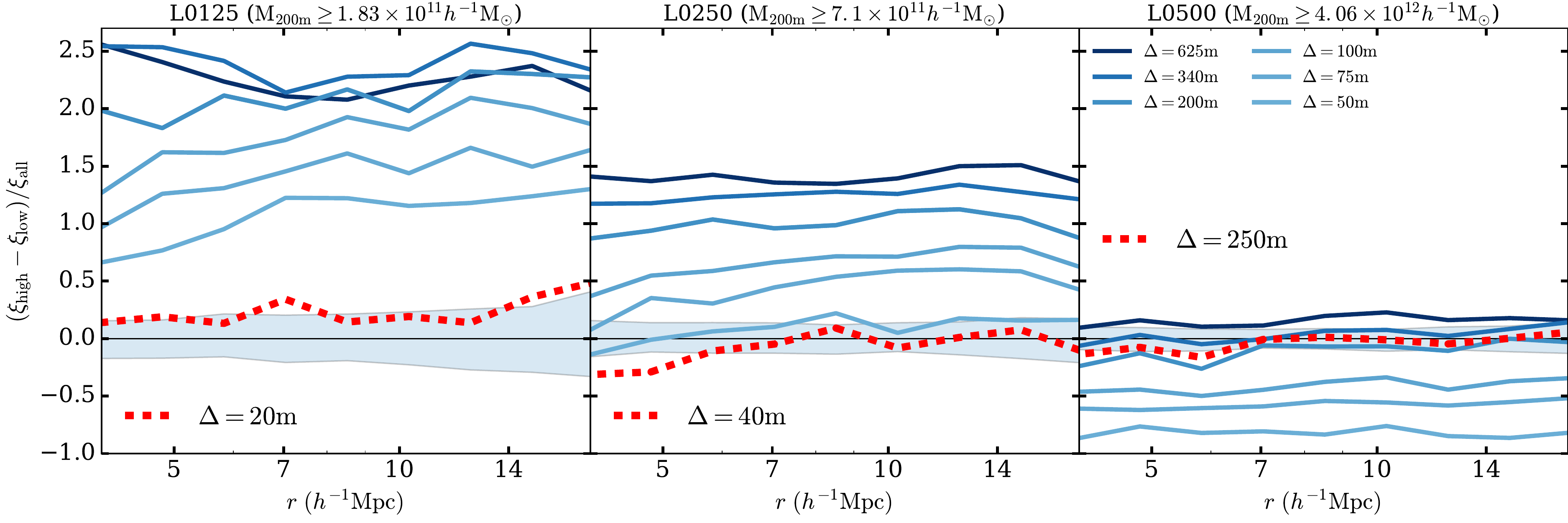}
	\caption{
Concentration-dependent correlation functions. In each panel, the solid lines plot the difference between the correlation function for the top 20\% and the bottom 20\% of haloes by NFW concentration, normalized by the correlation function of the entire halo sample. The lines correspond to different values of $\Delta$, with dark blue (light blue) corresponding to $\Delta = 625$m ($\Delta = 50$m). The red dashed line in each panel  corresponds to an overdensity that significantly reduces concentration-dependent halo clustering selected from the various values of halo overdensity that we explored. We will refer to these values throughout the text. The left (middle/right) panel shows the results for \simA \ (\simB /\simC) utilizing the low-mass (mid-mass/high-mass) halo threshold samples. To guide the reader, each panel is labelled by the minimum value of $M_{200{\rm m}}$ in the sample. Different values of $\Delta$ correspond to different mass thresholds as shown in Table~\ref{table:thresholds}. The shaded bands in each panel represent the level of statistical fluctuations due to finite sampling. In particular, the shaded bands contain $98\%$ of 200 CF ratios computed from randomly subsampling haloes in the $\Delta=200$ sample. In principle, each sample with a distinct $\Delta$ should have a distinct error band, but in practice they are all very similar to that of the $\Delta=200$m sample.
}
\label{fig:cc_cfcompare}
\end{figure*}
%-----------------------------------------------------------------------------------

We begin by studying the CFs of haloes in our mass threshold samples, sub-selected by auxiliary properties. As an example, Fig.~\ref{fig:cc_cfcompare} exhibits the difference between the clustering strengths of haloes in the top $20^{\mathrm{th}}$ percentile of NFW concentration compared to the haloes in the
bottom $20^{\mathrm{th}}$ percentile of NFW concentrations as a function of the overdensity parameter, $\Delta$, used to define the haloes. In order to scale out the gross scale dependence of the CFs, the two-point functions in Fig.~\ref{fig:cc_cfcompare} have been normalized by the clustering strength of the entire halo sample. 

If the clustering strength of haloes were independent of halo concentration, we would expect the lines in Fig.~\ref{fig:cc_cfcompare} to accumulate around zero (scattered about zero due to finite sample sizes). The evident deviations demonstrate that haloes of different NFW concentrations exhibit appreciably different clustering, a fact that is already well known and has been studied by a number of authors. Furthermore, it is clear that the strength and sign of assembly bias due to NFW concentration is 
strongly mass dependent for any fixed halo definition, a result that also agrees with the significant previous literature on halo assembly bias using conventional halo definitions.
At relatively low mass (the low-mass panel, $M_{200\text{m}} > 1.83 \times 10^{11} \hMsun$), high-concentration haloes are considerably more strongly clustered than low-concentration haloes, using the $\Delta = 200$m halo definition. At somewhat higher halo masses (e.g., the middle panel, $M_{200\text{m}} > 7.1 \times 10^{11} \hMsun$), this difference is markedly reduced. Finally, for the highest-mass haloes that we have the capability of studying (the right panel of Fig.~\ref{fig:cc_cfcompare}, $M_{200\text{m}} > 4.06 \times 10^{12} \hMsun$), the halo assembly bias effect is weaker and of opposite sign; low-concentration haloes are more strongly clustered than high-concentration haloes. 

%----------------- MCF CNFW
\begin{figure*}
	\centering
	\includegraphics[width=\textwidth]{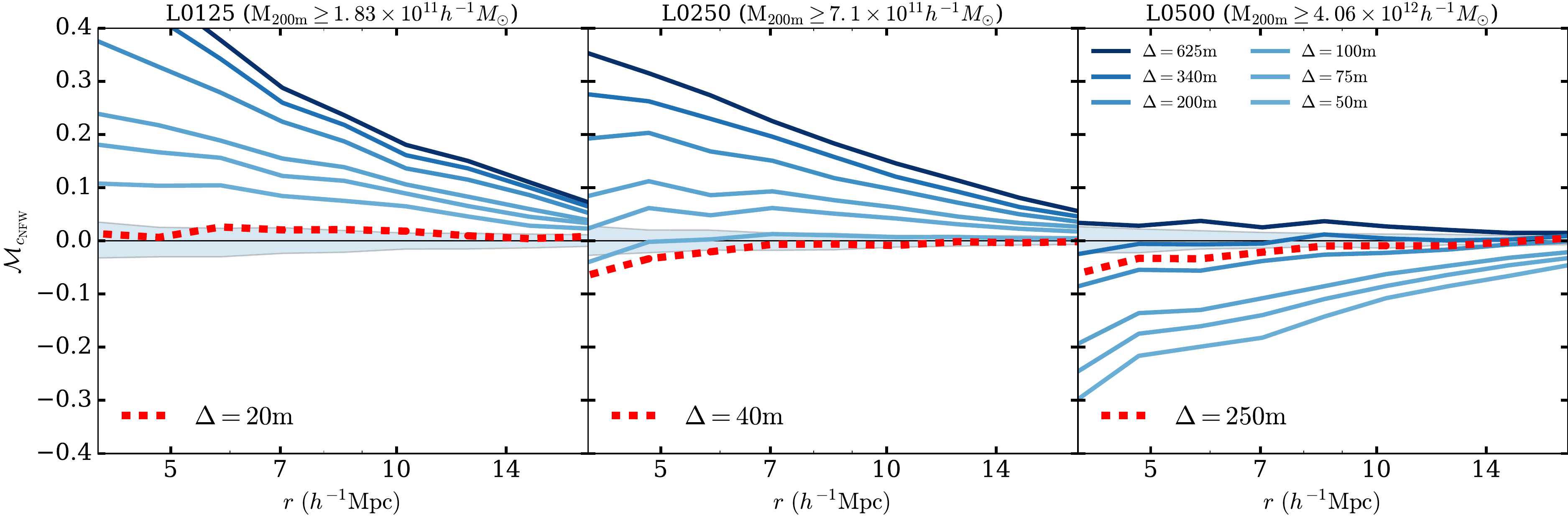}
	\caption{
The marked correlation function for the concentration defined according to the NFW profile, $c_{\rm NFW}$. The solid lines plot the marked correlation function using normalized ranks of NFW concentration as the mark. In each plot the lines correspond to different values of $\Delta$, with dark blue (light blue) corresponding to $\Delta = 625$m ($\Delta = 50$m). The red dashed lines correspond to the overdensities that greatly mitigate assembly bias for concentration (the same values of $\Delta$ as depicted by the red dashed lines in Fig.~\ref{fig:cc_cfcompare}). The top (middle/bottom) panel shows the results for the\simA \ (\simB /\simC) data set utilizing the low mass (mid mass/high mass) thresholds. To guide the reader, each panel is labelled by the minimum value of $M_{200{\rm m}}$ in the sample. Different values of $\Delta$ correspond to different mass thresholds as shown in Table~\ref{table:thresholds}. The shaded bands in each panel represent the level of statistical fluctuations due to finite sampling. In particular, the shaded bands contain $98\%$ of 200 MCFs computed from shuffling the halo marks among all of the haloes in the $\Delta=200$m sample. In principle, each sample with a distinct $\Delta$ should have a distinct error band, but in practice they are all very similar to that of the $\Delta=200$m sample.
}
\label{fig:cc_mcf_cnfw}
\end{figure*}
%------------------------------------

Another interesting effect is noticeable in the middle panel of Fig.~\ref{fig:cc_cfcompare}, corresponding to \simB~ and the mid-mass threshold. In this panel, the difference between the large-scale clustering of high- and low-concentration haloes is dramatically reduced for a halo definition with $\Delta \approx 40$m as compared to a more traditional halo definition, such as $\Delta=200$m. Further decreasing $\Delta$ leads to concentration-dependent clustering of opposite sense. Both the strength and the sense of halo assembly bias depend upon halo definition. This point may seem sensible in retrospect, but has not been addressed explicitly anywhere in the literature, despite its importance.

Comparing CFs across all three panels of Fig.~\ref{fig:cc_cfcompare}, it is clear that any specific conclusions about halo definitions that mitigate auxiliary property dependent clustering are mass dependent. For low-mass haloes (left panel), very low values of $\Delta \approx 20$m (and correspondingly large halo radii, as $R_{\Delta}$ varies roughly in proportion to $\Delta^{-1/3}$) are necessary in order to mitigate the concentration dependence of halo clustering. Yet, for higher-mass haloes (right panel), conventional values of $\Delta \sim 200-340$m yield little concentration-dependent clustering. In this case, decreasing $\Delta$ (increasing $R_{\Delta}$) results in significantly {\em increased} concentration dependent halo clustering. The reasons for 
these changes are of interest and we return to the interpretation of these results below.

Notice that in all panels of Fig.~\ref{fig:cc_cfcompare}, the effect of concentration-dependent clustering is mildly scale-dependent. Moreover, this scale dependence is evident for all values of $\Delta$. Simply defining haloes with a different value of $\Delta$ does not suffice to eliminate concentration-dependent clustering to high precision on all of the scales that we study. In this discussion and throughout, we focus primarily on the large scale clustering, which we take to mean clustering on scales significantly larger than the radii, $R_{\Delta}$ of the haloes in our samples. In the language of halo occupation models of galaxy clustering, we focus on two-halo clustering scales.

The broad conclusions that we draw from examining CFs, such as those in Fig.~\ref{fig:cc_cfcompare} are robust to the choices we have made. For instance, we would draw the same conclusions by examining other halo subsamples (e.g., the top and bottom $10^{\rm{th}}$ percentile of haloes by concentration). Furthermore, examining $c_{\mathrm{V}}$ rather than $c_{\mathrm{NFW}}$ also does not alter our general conclusions. Therefore, we do not show these additional results in the interest of brevity. More generally, we find that for each of the halo properties that we have studied, the conclusions drawn from examining CFs are very similar to those drawn from studying MCFs. Consequently, we now move on to a more comprehensive discussion of the strength of auxiliary property-dependent halo clustering using MCFs. 

%-----------------------------------------------------------------
\subsection{Mark Correlation Functions}
\label{sub:mcfresults}

%------------------------------------- MCF CV
\begin{figure*}
	\centering
	\includegraphics[width=\textwidth]{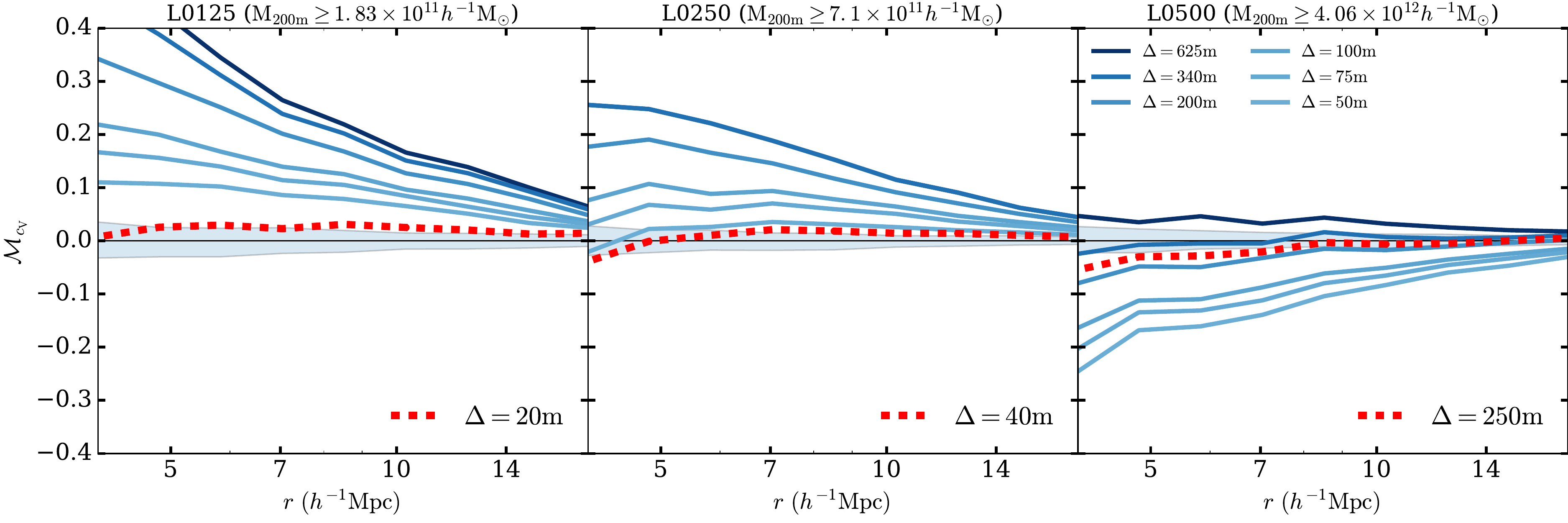}
	\caption{	
The same as Fig.~\ref{fig:cc_mcf_cnfw} for the mark of $c_{\rm V}$.
Red dashed lines indicate the $\Delta$ values that remove assembly bias from \emph{concentration}. 
}
	\label{fig:cc_mcf_cV}
\end{figure*}
%---------------------------------------------------

%------------------------
\subsubsection{Halo Concentration}

The NFW concentration, $c_{\mathrm{NFW}}$, MCF is shown in Fig.~\ref{fig:cc_mcf_cnfw}. The shaded bands in the figure delineate the statistical fluctuations in MCFs induced by finite sampling as discussed in Section~\ref{subsection:clusteringstatistics}. In principle, each sample should be compared with a distinct band because the samples do not contain the same objects and, consequently, may exhibit different levels of statistical fluctuation. In practice, these error bands are very similar across all samples. Therefore, in Fig.~\ref{fig:cc_mcf_cnfw} and the similar plots that follow, we show only the error bands that correspond to the $\Delta=200$m halo samples. The error bands for the samples selected according to different values of the overdensity parameter are very similar and serve only to obscure the information in the figure. 

Qualitatively, Fig.~\ref{fig:cc_mcf_cnfw} exhibits the same features that are evident in Fig.~\ref{fig:cc_cfcompare}: more concentrated haloes are significantly more clustered in the low-mass, L0125 halo sample, and concentration-dependent halo clustering weakens and reverses sense as halo mass increases (at fixed $\Delta$). This is consistent with previous work on assembly bias \citep[e.g.,][]{wechsler_etal06,sunayama_etal16}. For the mid-mass cut L0250 sample with $\Delta=40$, the large-scale concentration dependence of halo clustering has been reduced so as to be consistent with zero within the statistical limitations of the simulation. 

Fig.~\ref{fig:cc_mcf_cV} is a similar plot for the MCF of the velocity-ratio concentration, $c_{\mathrm{V}}$. This figure exhibits qualitatively very similar features to Fig.~\ref{fig:cc_mcf_cnfw}, a fact that is not surprising given that we already know that $c_{\mathrm{NFW}}$ and $c_{\mathrm{V}}$ quantify largely redundant information about their haloes. This demonstrates that our results regarding the halo boundary dependence (or $\Delta$ dependence) of halo clustering is not driven by details of a fit of the density profile to the NFW functional form, which is an increasingly poor description of the halo profiles for halo-centric distances significantly larger than $R_{200\mathrm{m}}$ \citep[e.g.,][]{diemer_kravtsov14} 
\footnote{We provide a further demonstration that the decrease in assembly bias due to the decrease in $\Delta$ in this case is not due to the fidelity with which the NFW profile describes halo density profiles at large radii in \S~\ref{section:halodefmit}.}.

%----------------------
\subsubsection{Halo Shape}

\begin{figure*}
	\centering
	\includegraphics[width=\textwidth]{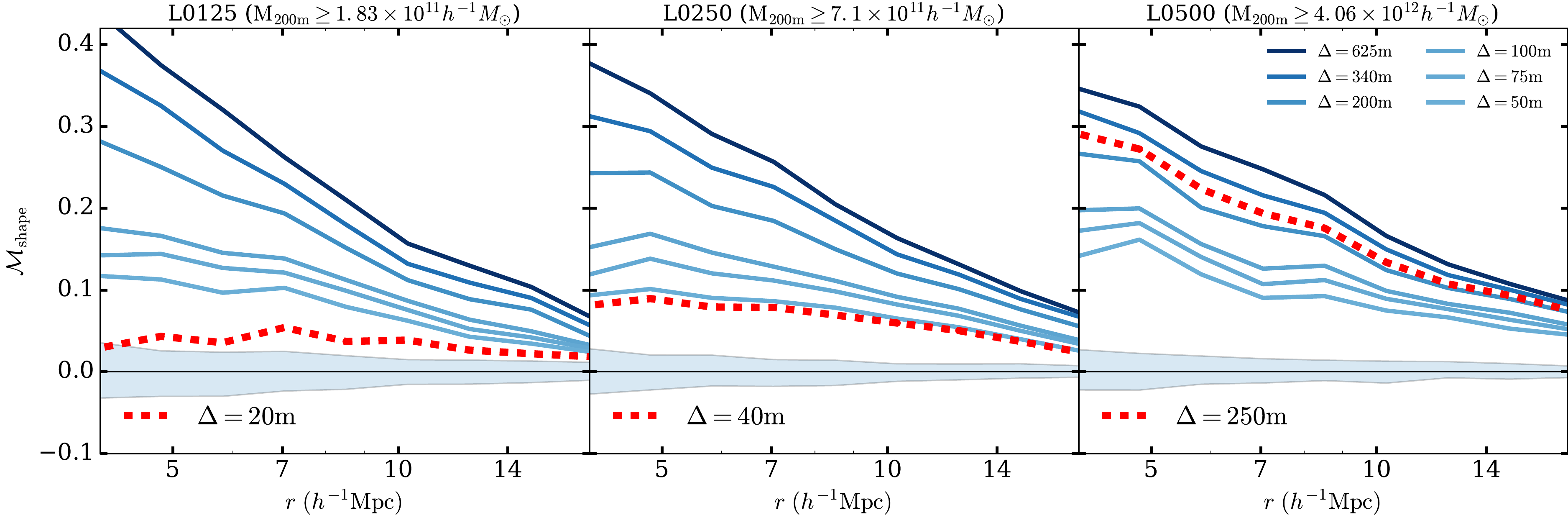}
	\caption{
The same as Fig.~\ref{fig:cc_mcf_cnfw} for a halo shape MCF.
Red dashed lines indicate the $\Delta$ values that remove assembly bias from \emph{concentration}. 
}
	\label{fig:cc_mcf_s}
\end{figure*}

Moving on from concentrations, Fig.~\ref{fig:cc_mcf_s} illustrates MCFs in which the mark is the normalized rank of the shape parameter, $s$, of the halo. For the entire range of halo masses studied, the fiducial halo definition shows that more spherical haloes (those with {\em larger} marks, that is {\em larger} values of $s$) are more strongly clustered. Furthermore, Fig.~\ref{fig:cc_mcf_s} shows that increasing the sizes of haloes (decreasing $\Delta$, thereby increasing the radius of the halo boundary) reduces this shape-dependent halo clustering. Unlike the case of concentration, shape-dependent halo clustering does not have a strong dependence upon halo mass.

%--------------------------------------------------- Shape MCFs
%--------------------------------------------

Shape-dependent halo clustering also behaves differently from concentration-dependent halo clustering in that it persists even if haloes are defined relative to an extremely low overdensity parameter, such as $\Delta=20$. In no case in Fig.~\ref{fig:cc_mcf_s} is shape-dependent clustering consistent with zero. No reasonable value of $\Delta$ seems to be capable of removing the enhanced clustering at the scales that we study, though the magnitude of shape-dependent clustering can be reduced by decreasing $\Delta$ (and increasing halo radii). In particular, the most effective values of $\Delta$ for removing concentration-dependent clustering, the red-dashed lines in Fig.~\ref{fig:cc_mcf_s}, do not correspond with removing assembly bias from shape. 

%---------------- spin MCF
\begin{figure*}
	\centering
	\includegraphics[width=\textwidth]{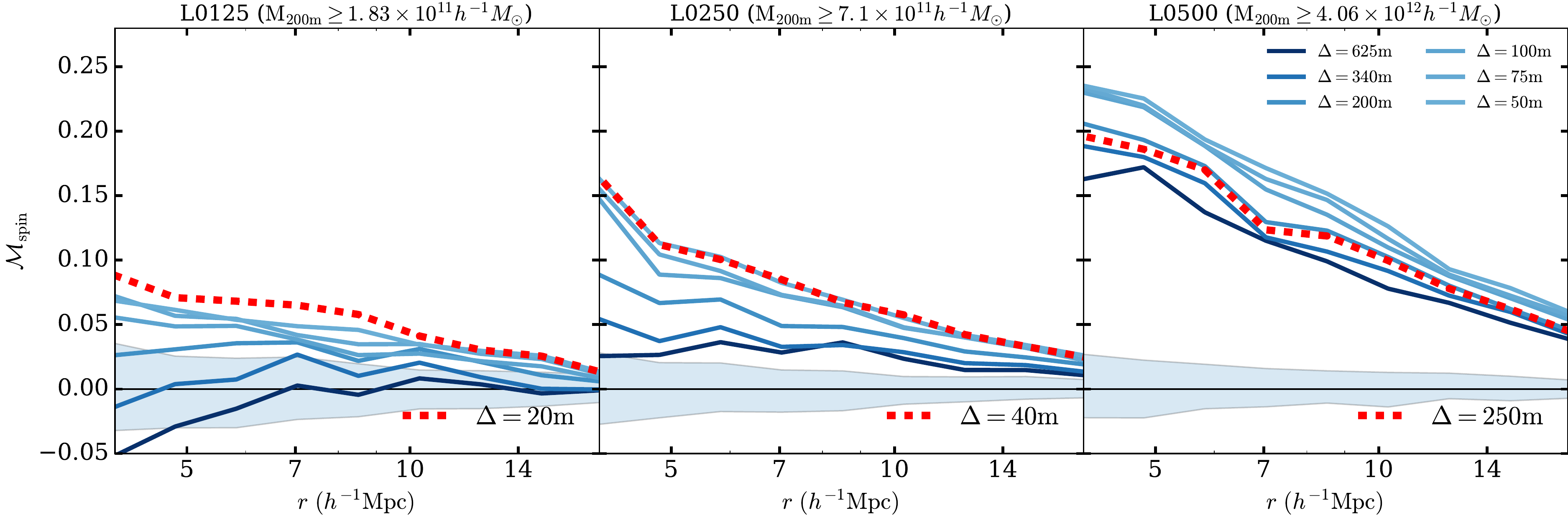}
	\caption{
The same as Fig.~\ref{fig:cc_mcf_cnfw} for a halo spin MCF.
Red dashed lines indicate the $\Delta$ values that remove assembly bias from \emph{concentration}. 
}
	\label{fig:cc_mcf_spin}
\end{figure*}
%-------------------------------

%------------------
\subsubsection{Halo Spin}

The spin MCFs are shown in Fig.~\ref{fig:cc_mcf_spin}. In most cases, haloes that are more clustered show higher values for spin than would be expected of a random sample from the one-point spin distribution, a result that is consistent with the previous literature on halo spin-dependent clustering.
In detail, spin dependent halo clustering is mass dependent. In our low-mass sample ($1.8 \times 10^{11}\, \hMsun \le M_{200m} < 7.1 \times 10^{11}\, \hMsun$), haloes defined by overdensities of $\Delta=625$m (corresponding to $200$c) to $\Delta=340$m exhibit little spin-dependent clustering in excess of that expected from finite sampling of the spin distribution. It is also evident from Fig.~\ref{fig:cc_mcf_spin} that spin-dependent clustering increases in strength with halo mass. In the high-mass sample, spin-dependent halo clustering is a strong effect on all scales for all halo definitions. While a conventional halo definition, such as $\Delta \sim 340{\rm m}-625{\rm m}$, would yield minimal spin-dependent halo clustering, at higher masses significantly larger values of $\Delta$ would be needed to mitigate spin assembly bias.

%----------------------------------------
\begin{figure*}
	\centering
	\includegraphics[width=\textwidth]{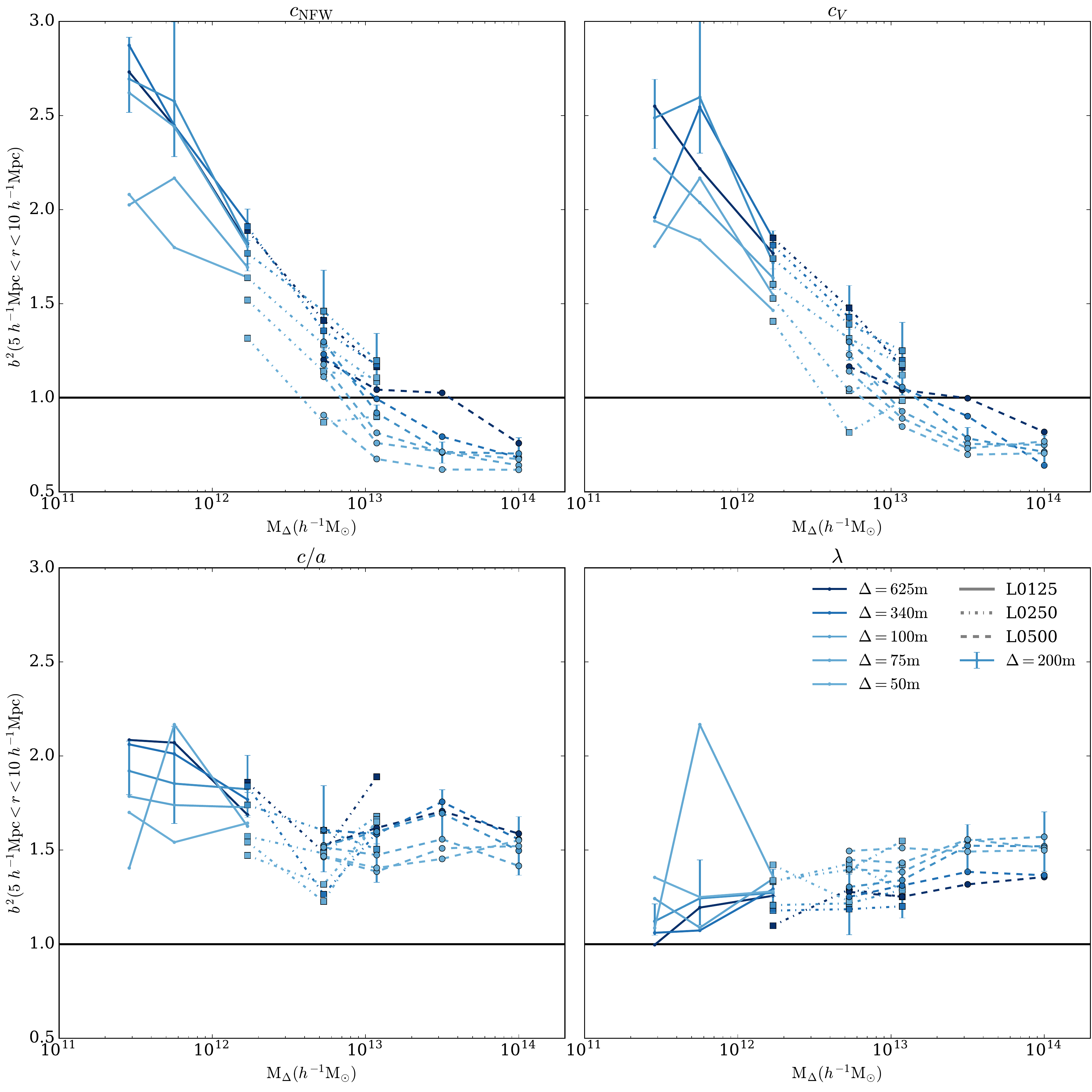}
	\caption{
The relative clustering bias of haloes as a function of halo mass for various halo auxiliary properties. For illustration in this plot, we show the clustering strength of the highest 20-percentile according to each halo property relative to the population of all haloes. The properties of interest are labelled at the top of each panel. Within each panel, we show clustering biases for six values of $\Delta$. The dark blue (light blue) line uses a halo definition drawn from $\Delta = 625$m ($\Delta = 50$m). The solid (dot-dashed, dashed) lines use host haloes from the \simA{} (\simB,\simC) catalogues. The error bars on the $\Delta=200$ samples are similar to the errors from other samples, which are not shown for clarity.
}
\label{fig:biascompare}
\end{figure*}
%--------------------------------------------

As with halo shape, it can be seen that the values of overdensity that best mitigate concentration-dependent halo assembly bias (the red-dashed lines) do not serve to mitigate assembly bias with respect to halo spin. Indeed, halo spin exhibits a response that is qualitatively different from concentration as $\Delta$ is varied. Increasing halo radii by decreasing $\Delta$ only serves to increase the magnitude of spin-dependent assembly bias; the newly-defined haloes are even more likely to have high spin if they are found in a relatively close pair with another halo. We speculate that this may be caused by material on the outskirts of haloes that can carry large amounts of angular momentum with respect to the halo centre (due to the large value of the impact parameter to the halo centre) and that a significant amount of material contributing to angular momenta at this radii may be indicative of an overdense large-scale environment. Follow up work is necessary to determine whether or not this speculation is correct and we intend to explore this in a forthcoming paper.

%--- Summary of Halo Clustering
\subsection{Summary of Halo Clustering as a Function of Halo Definition}

Various forms of assembly bias, and their mass dependences, have already been identified in the literature. Our work confirms these previous results for halo concentrations \citep{wechsler_etal06,faltenbacher_white10,sunayama_etal16}, halo shapes \citep{bett_etal07,hahn_etal07b,hahn_etal07b,faltenbacher_white10,lacerna_padilla12,vandaalen_etal12}, and halo angular momenta  \citep{bett_etal07,hahn_etal07a,hahn_etal07b,lacerna_padilla12}. Our work emphasizes the fact that auxiliary property dependent halo clustering, which we refer to as assembly bias, is strongly dependent upon halo definition. 

Fig.~\ref{fig:biascompare} summarizes these aspects of auxiliary property dependent clustering for each of the properties we study. Fig.~\ref{fig:biascompare} shows the {\em relative} halo clustering bias of high-$p$ haloes, where $p$ is the property of interest, compared to all haloes defined as: 
\beq
b^2(r) = \xi_{p,80\%} / \xi_{\mathrm{all}},
\eeq
where $\xi_{p,80\%}$ designates the clustering of haloes in the $80^{\mathrm{th}}$ percentile of $p$ at fixed mass (that is, the haloes with the $20\%$ highest $p$ values). Though quantitatively different, choosing percentiles other than $80\%$ lead to similar qualitative conclusions. To summarize clustering strength with a single relative bias, we computed the ratio of the correlation functions averaged over a wide bin of separations from $r = 5-10\, \hMpc$. The wide bin width is justified partly by the mild scale dependence shown in Fig.~\ref{fig:cc_cfcompare}. We estimated the errors on the relative bias by recomputing $\xi_{\mathrm{all}}$ using randomly-selected subsamples of equal number to the subsample used to compute $\xi_{p,80\%}$, that is, using one fifth of all haloes. The error bars show the $68\%$ range, centred on the median, about which the values of $b^2(r)$ lie. 

Each panel of Fig.~\ref{fig:biascompare} allows us to focus on the mass dependence of halo clustering and the dependence on halo definition. Focus first on the top two panels, which deal with our two halo concentration measurements, NFW concentration and velocity ratio concentration. Both measures of concentration exhibit a strong mass dependence to assembly bias. Haloes at high masses ($M_{\Delta} \ge 10^{13}\, \hMsun$) exhibit assembly bias that is quite distinct from that of low mass ($M_{\Delta} \le 10\times10^{12}\, \hMsun$) haloes. 
At high masses, high-concentration haloes are less strongly clustered than low-concentration haloes. At low-masses the sense of assembly bias is reversed and high-concentration haloes cluster more strongly. Notice also that at high masses, concentration-dependent halo clustering is only a modest function of halo mass.

Concentration dependent halo clustering has a significant dependence upon halo definition. In particular, the strength of clustering of the haloes above the $80^{\rm{th}}$ percentile in concentration varies by as much as $\sim 50\%$ among the overdensities that we have investigated. For all halo definitions, concentration-dependent clustering exhibits a similar trend with halo mass. For no value of $\Delta$ does the trend with mass become significantly less prominent, suggesting that a simple mass-independent re-definition of halo boundaries alone cannot eliminate halo assembly bias. Concentration-dependent clustering changes sense at a mass that varies by an order of magnitude, from $\sim 3 \times 10^{12}\, \hMsun$ to $\sim 3 \times 10^{13}\, \hMsun$, as $\Delta$ varies from $\Delta=50$ to $\Delta=625$. These significant variations in the strength of concentration-dependent clustering as a function of halo definition suggest that significant care must be taken in comparisons of various results in the extant literature.

The bottom two panels of Fig.~\ref{fig:biascompare} demonstrate that auxiliary property dependent clustering can behave in a markedly different manner depending upon the halo property under consideration. These two panels show shape (left) and spin (right) dependent clustering as a function of halo mass and halo definition ($\Delta$). Each demonstrates halo assembly bias that is only weakly dependent on halo mass in comparison to concentration-dependent clustering. In the lower-left panel, note that changing to smaller values of $\Delta$ results in reduced assembly bias, though no definition explored was sufficient to remove assembly bias entirely. 

It is worth emphasizing that the $\Delta$ dependence of assembly bias in Fig.~\ref{fig:biascompare} is non-trivial. At first glance, the reader may be tempted to think that each of the lines in any individual panel of Fig.~\ref{fig:biascompare} could be made to overlap by plotting the bias with respect to a common mass scale for each halo, rather than for $M_{\Delta}$ for each value of $\Delta$. In other words, the reader may be tempted to think that the only reason that the lines in Fig.~\ref{fig:biascompare} do not overlap is trivially due to the shift in halo masses caused different $\Delta$ (e.g., Fig.~\ref{fig:deltacompare}). However, this is {\em not} the case. Consider the panels of Fig.~\ref{fig:biascompare} depicting halo concentration. Suppose that we had chosen to plot halo relative bias as a function of the independent variable $M_{200m}$, the $\Delta=200$m mass of each halo. For any individual object, if $\Delta' < \Delta$, then $M_{\Delta'} > M_{\Delta}$. Consequently, for any individual halo $M_{200{\rm m}} > M_{625{\rm m}}$ while $M_{200{\rm m}} < M_{50{\rm m}}$. Therefore, representing the relative halo bias on a common mass scale (e.g., $M_{200{\rm m}}$) for all haloes moves the curves in Fig.~\ref{fig:biascompare} {\it further away from one another}. The analogous statement is true for all panels of Fig.~\ref{fig:biascompare}. The point is profound because it demonstrates that the $\Delta$ dependence of assembly bias is {\em not} caused simply by the shift in mass scale, but rather by the selection of host haloes to include in the halo sample. 

%---------------------------------------------------------------------------------------------
\section{Discussion}
\label{section:discussion}
%---------------------------------------------------------------------------------------------

%----------- NFW match figure
\begin{figure*}
	\centering
	\includegraphics[width=\textwidth]{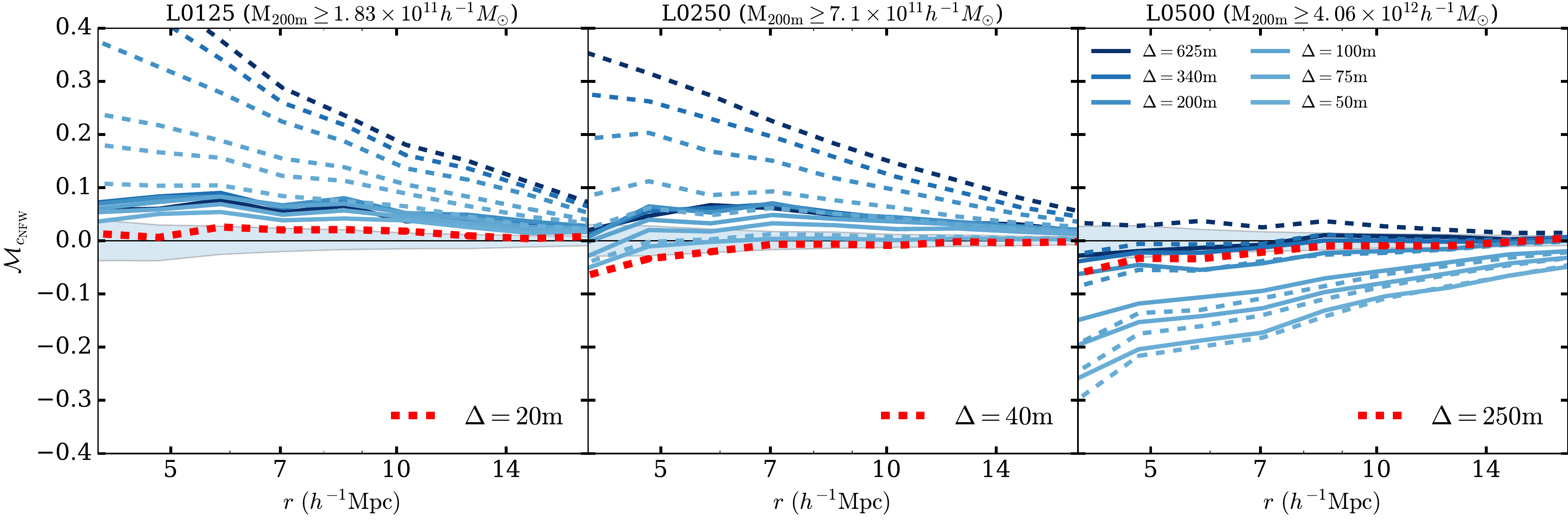}
	\caption{The marked correlation function for the NFW-defined halo concentration parameter matched to baseline halo catalogues. The dashed lines reproduce the MCFs of Fig.~\ref{fig:cc_mcf_cnfw} for reference. The solid lines plot the marked correlation function using NFW-defined halo concentration as the mark, for a catalogue of host haloes matched to the host haloes in the baseline catalogue corresponding to the $\Delta$ that best mitigates assembly bias (shown as a red dashed line in each panel). For each value of $\Delta$ only host haloes that also appear in the baseline $\Delta$ halo catalogue are used in constructing the MCF shown by the solid lines. For each value of $\Delta$, halo properties are determined using the indicated value of $\Delta$ and {\em not} the baseline value. The baseline $\Delta$ only determines which haloes are included in the sample. See the main text for further details on sample selection. All lines correspond to different values of $\Delta$, with dark blue (light blue) corresponding to $\Delta = 625$m ($\Delta = 50$m). The left (middle/right) panel shows the results for the \simA \ (\simB /\simC) data set utilizing the low mass (mid mass/high mass) cutoffs. The shaded bands represent 2-sigma confidence regions generated by randomization of the marks. This figure demonstrates that most of the assembly bias is removed by the classification of host and satellite haloes, rather than by altering the measurement of halo properties such as concentration. 
\label{fig:hvm_mcf_cnfw}
}
\end{figure*}
%-----------------------------------------------------

%----------------------------
\subsection{Broad Conclusions}
%----------------------------

We have studied the clustering of dark matter haloes as a function of halo properties other than mass. We have confirmed that for conventional halo definitions halo clustering strength is a strong function of the  ``auxiliary properties'' that we studied, namely halo concentration (either measured through a fit to the NFW profile or assigned non-parametrically as the ratio of the maximum circular velocity to the virial velocity), halo shape, and halo spin. These findings are consistent with the now significant literature on the subject of halo assembly bias. \citep[e.g.,][]{peacock_smith00, wechsler_etal02,sheth_tormen04, gao_etal05, zentner_etal05, allgood_etal06, harker_etal06, wechsler_etal06, croton_etal07, zentner07, dalal_etal08, lacerna_padilla12, zentner_etal14, mao_etal15, sunayama_etal16}

The goal of this study has been to explore auxiliary property dependent halo clustering as a function of halo definition, parametrized by overdensity parameter $\Delta$. This exploration was motivated, in part, to determine whether alternative halo definitions can mitigate the dependence of halo clustering on these ``auxiliary properties.'' We find that these alternative halo definitions do {\em not} generally mitigate the effects of assembly bias. We are led to this conclusion for two reasons. First, no single halo definition serves to mitigate auxiliary property dependent clustering for all of the auxiliary properties that we have investigated. Indeed, modified halo definitions, with low values of $\Delta$, may lead to stronger auxiliary property dependent clustering as is the case with halo spin, for example.

Second, the halo definitions that mitigate assembly bias are strongly halo mass dependent. In the case of halo concentration, which we examine in the most detail, assembly bias is greatly mitigated for halo definitions of $\Delta \approx 20$m, $\Delta \approx 40$m, and $\Delta \approx 250$m for haloes in our $M \geq 1.83 \times 10^{11} \hMsun$, $M \geq 7.1 \times 10^{11} \hMsun$, and $M \geq 4.06 \times 10^{12} \hMsun$ mass samples respectively. Of course, this mass dependence may not be surprising given the well-known mass dependence of assembly bias. Physically, this is likely due to the fact that concentration-driven assembly bias may be caused by distinct mechanisms at the low-mass \citep[e.g., see][]{wang_etal07,dalal_etal08,warnick_etal08,hahn_etal09,ludlow_etal09,sunayama_etal16} and high-mass \citep[e.g.,][]{zentner07,dalal_etal08} ends of the halo mass spectrum. 

These two observations together suggest that it may well be possible to mitigate assembly bias for a single halo property (e.g., concentration) over a finite range of halo masses by using a mass-dependent halo definition. However, incorporating this result into self-consistent halo identification algorithms that are valid over a wide range of halo masses would require more complex, likely iterative, halo finding process. Furthermore, a simple mass-dependent halo definition may not be able to mitigate assembly bias. For example, consider the fact that assembly bias at low mass is, at least in part, mitigated by re-classifying harassed haloes in the environments of significantly more massive neighbour haloes as subhaloes of these neighbours. It may be possible that a halo-by-halo definition scheme, perhaps according to halo splashback radii \citep{diemer_kravtsov14,diemer_kravtsov15,mansfield_etal16,diemer_etal17}, may better mitigate assembly bias. We discuss the particular case of concentration-dependent clustering and the relationship to splashback radii further below.

Our work also emphasizes the important dependence of halo assembly bias upon halo definition, which is not yet widely appreciated. In each of Figs. \ref{fig:cc_mcf_cnfw} through \ref{fig:cc_mcf_spin}, the strength of assembly bias is a strong function of halo definition. This definition dependence is not restricted to extreme choices of the overdensity parameter $\Delta$. The difference in the strength of assembly bias between a ``virial'' halo definition, with $\Delta=340$m, and a definition with $\Delta=200$m is considerable. Likewise, defining haloes using an overdensity of 200 relative to the critical density, corresponding roughly to $\Delta \approx 625$m in our notation, also leads to significant, quantitative differences in the strength of assembly bias. Indeed, distinct halo definitions may even lead to assembly bias of opposite sense. For example, at fixed halo mass, higher concentration haloes may be more strongly clustered by one halo definition and more weakly clustered by another. It is interesting to consider that some of the variety of results pertaining to the strength of assembly bias in the literature may be partly induced by the different halo definitions used by different authors.

%-------------------------------------------
\begin{figure*}
	\centering
	\includegraphics[width=\textwidth]{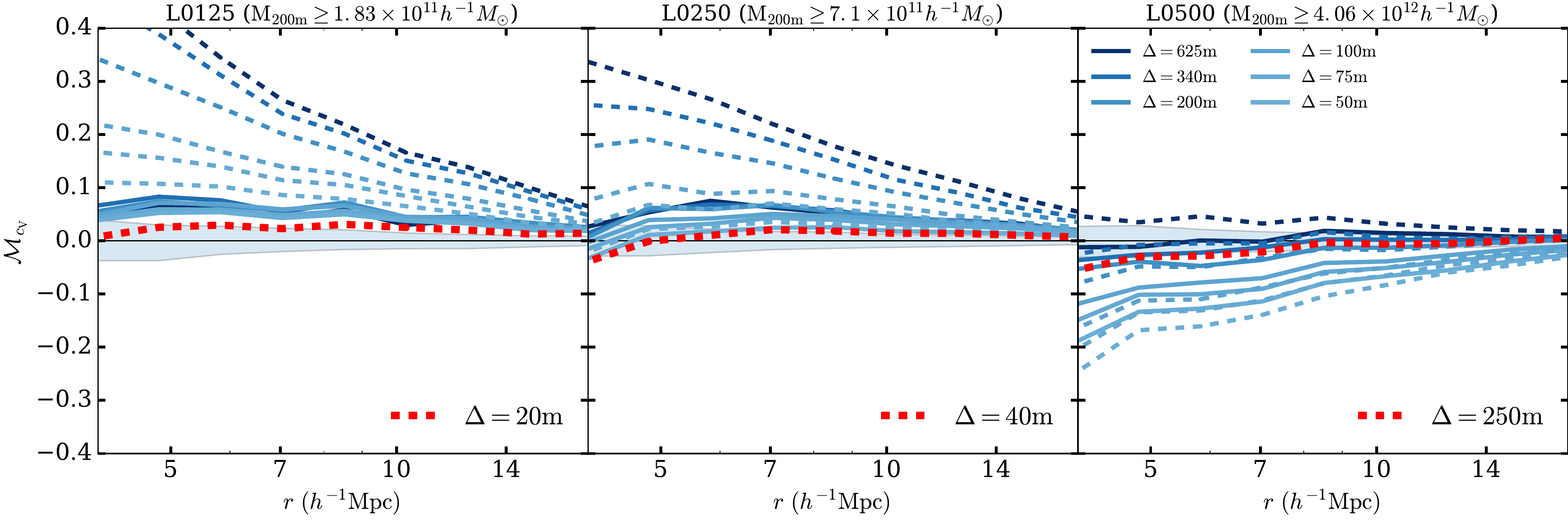}
	\caption{As Fig.~\ref{fig:hvm_mcf_cnfw}, with the mark defined as velocity ratio defined concentration.
}
	\label{fig:hvm_mcf_cV}
\end{figure*}
%-----------------------------------------------

Aside from previous work on halo splashback radii (which we discuss in more detail in Section~\ref{subsection:splashback}), the work that is most closely related to ours is that of \citet[][hereafter LP11]{lacerna_padilla11}. Motivated by previous work to explain age-related assembly bias \citep{wang_etal07,wang_etal08,warnick_etal08,dalal_etal08,hahn_etal09}, LP11 considered defining halos by alternative means. There are important distinctions between the work of LP11 and our work. LP11 developed a parametrized prescription for the mass of a halo in which parameters were fit to remove assembly bias as defined by the ages of the galaxies within the halos according to the semi-analytic galaxy formation model of \citet{lagos_etal08}. Our work relates to the halos themselves and is an exploration of the assembly bias of halos, with no reference to the galaxies that may reside within them. This is a different perspective that is intended to empower empirical models of galaxy clustering that can be used to interpret galaxy clustering data. Furthermore, the halo properties we study, namely concentration, spin and shape, are quite distinct from the galaxy ages studies by LP11. As a final, technical point, LP11 did not repeat the identification of halos within their simulations self-consistently for each new halo definition. We have self-consistently repeated the halo identification process (using {\tt ROCKSTAR}) for each halo definition we have explored. We have not considered halo age explicitly because repeating the calculation of the halo merger tree self-consistently for each halo definition is computationally demanding. We will explore halo age in a forthcoming manuscript.

%------------------------------------------------
\subsection{How Does Halo Definition Mitigate Concentration-Dependent Halo Clustering}
\label{section:halodefmit}
%------------------------------------------------

As mentioned above, our results indicate that one may mitigate some auxiliary property dependent clustering through a mass-dependent choice of halo definition. It is interesting to ask why halo redefinitions may be helpful. In this subsection, we discuss the particular case of halo concentration dependent clustering in more detail in part because concentrations may be the most physically interesting property that we have studied.

Halo redefinitions may mitigate concentration-dependent halo bias for at least two reasons. The first reason is the physical motivation for exploring alternative halo definitions. In particular, it may be that alternative halo definitions (presumably with smaller $\Delta$) provide a more effective grouping of objects that have been strongly affected by interactions into one single halo as  illustrated in the cartoon in Fig.~\ref{fig:halocartoon}. If this is the case, alternative halo definitions may offer a more practical separation between the linear and highly nonlinear regimes and may represent a pragmatic step forward. However, alternative halo definitions may also reduce auxiliary property dependent clustering through a second mechanism. To be specific, it is possible that the details of measuring halo properties using these new halo definitions introduce new sources of noise into the measurements. The inherently noisier measurements then lead to reduced correlations. In this second case, the reduction in correlations simply arises because the property of interest may be {\em less} informative about the halo itself.

In the case of halo concentration, noise may be introduced in numerous ways. For example, the NFW concentration $c_{\mathrm{NFW}}$ is determined by a fit to the NFW profile. Inferred values of $c_{\mathrm{NFW}}$ will depend upon the degree to which the density profiles of the haloes follow the NFW functional form within some radius $R_{\Delta}$ that is different from traditional halo radii, such as $\sim R_{200\text{m}}$. At large halocentric distances ($r \gtrsim R_{200\text{m}}$) halo profiles are known to deviate markedly from the NFW form. It may be possible to reduce assembly bias by redefinitions if one probes scales on which haloes deviate from NFW in a way that is not well correlated with the interior structure of the halo (particularly the location of the NFW scale radius); however, such a reduction in assembly bias is of limited practicality because it arises from characterizing a halo by a quantity that is less informative about its interior structure. Of course, it is worth noting that the velocity-defined concentration $c_{\rm V}$ is a non-parametric measure of concentration and should be less subject to such effects. 

We can explore in more detail the degree to which the mitigation of environmental effects by halo redefinition is due to the introduction of noise that is uncorrelated with environment into the measurement of halo properties. We describe one method for doing this in the remainder of this subsection. The cartoon of Fig.~\ref{fig:halocartoon} may be a useful reference for the reader. 

Consider first that all host haloes that are present in the halo catalogues constructed from any one specific value of the overdensity threshold (e.g., $\Delta=40$m) are also present as host haloes in the halo catalogues constructed with all higher values of threshold density (e.g., $\Delta=200$m and, in fact, all $\Delta \ge 40$m). The converse is not true because lowering $\Delta$ increases halo radii so that host haloes at higher values of $\Delta$ may become subhaloes at lower values of $\Delta$\footnote{Indeed, this is largely the motivation for exploring various $\Delta$.}. This reclassification as subhaloes is the fate of haloes C and D in Fig.~\ref{fig:halocartoon} as the overdensity threshold is decreased from $\Delta=200$m to $\Delta=20$m.

We consider matching haloes across different catalogues, constructed using different values of $\Delta$. For this exercise, we match haloes to a {\em baseline} catalogue corresponding to a value of $\Delta$ that best mitigates concentration-dependent clustering at each mass bin. These baseline values of $\Delta$ are those delineated by the dashed, red lines in Fig.~\ref{fig:cc_cfcompare}--\ref{fig:cc_mcf_spin}. In this exercise, we consider the clustering of only those haloes that are classified as host haloes in the baseline catalog. 

Specific examples may help to clarify our procedure. Consider, for example, the low-mass sample constructed from the \simA{} simulation. The baseline overdensity in this case is $\Delta=20$m. So, for the low-mass sample, we study the clustering of haloes designated as host haloes according to a $\Delta=20$m halo definition. However, we assign these haloes properties according to their definitions using other values of $\Delta$. For example, we may study the clustering of these haloes as a function of the masses and concentrations defined using a $\Delta=200$m halo definition. Referring back to Fig.~\ref{fig:halocartoon}, we study the clustering of haloes A and E using halo properties derived from the particles within $R_{200{\mathrm{m}}}$. This is useful because it preserves the original halo properties defined using a conventional halo definition (such as $\Delta=200$m), but it removes those haloes from consideration that we may expect to be altered by interactions with neighbouring haloes. In the cartoon of Fig.~\ref{fig:halocartoon}, we are not considering the clustering of haloes C and D. They are subhaloes in the $\Delta=20$m halo catalog. 

A second example may be useful and is relevant to the high-mass sample that we study, constructed from the \simC{} simulation. In this case, the baseline overdensity is $\Delta=250$m. We consider the clustering of all objects that are host haloes in the $\Delta=250$m halo catalog. If we define properties relative to any $\Delta \ge 250$m, then this is very much the same as in the previous example in that using the baseline catalogue serves to eliminate some haloes from consideration. However, for $\Delta < 250$m, some of the objects that we consider are subhaloes. In the context of Fig.~\ref{fig:halocartoon}, this is analogous to studying the clustering of haloes A, C, D, and E using properties defined by all particles within the dashed boundaries. They key point in all of these examples, is that the baseline halo catalogue is used to define the sample of haloes whose clustering we study.

Figure~\ref{fig:hvm_mcf_cnfw} shows the results of this investigation. As a reference for the reader, the dashed lines in Fig.~\ref{fig:hvm_mcf_cnfw} reproduce the same MCFs depicted in Fig.~\ref{fig:cc_mcf_cnfw}. The solid lines in Fig.~\ref{fig:hvm_mcf_cnfw} show MCFs using the samples matched to the baseline halo catalogues. To be explicit, in the left panel of Fig.~\ref{fig:hvm_mcf_cnfw} we show the MCFs of all haloes that are host haloes in the  $\Delta=20$m baseline halo catalogue for which the mass and concentration (the mark in this case) of each halo has been computed using the value of $\Delta$ that corresponds to the colour of the curve. The analogous statement is true for the middle and right panels. For both the solid and dashed dark blue lines, corresponding to $\Delta=625$m, the haloes are assigned masses and concentrations using a $\Delta=625$m. In the case of the solid line, only haloes defined as hosts in the $\Delta=20$m baseline catalogue are included in the computation of the MCF. For each halo sample, the baseline values of $\Delta$ are chosen effectively to remove assembly bias at large scales based on the results in Fig.~\ref{fig:cc_mcf_cnfw} and Fig.~\ref{fig:cc_mcf_cV}. These baseline overdensities are $\Delta=20$m for the \simA{} sample, $\Delta=40$m for the \simB{} sample, and $\Delta=250$m for the \simC{} sample. 

Compare pairs of dashed and solid lines at the same $\Delta$ threshold (same colour) in Fig.~\ref{fig:hvm_mcf_cnfw}. The difference between a pair of solid and dashed lines at fixed $\Delta$ is caused entirely by the exclusion of some haloes from the lower $\Delta$ catalogue due to a change in halo definition. For the low-mass sample (\simA) and mid-mass sample (\simB), the solid lines exhibit greatly reduced concentration-dependent clustering. We conclude that for relatively low halo masses ($M_{200{\rm m}} \lesssim 4 \times 10^{12}\, \hMsun$), selecting particular haloes as hosts eliminates the majority of concentration-dependent clustering. In other words, the majority of the assembly bias signal in this mass range is caused by haloes that are nearby neighbours of other, larger haloes. At high mass (the \simC{} sample), the story is somewhat different. The reclassification of haloes as hosts or satellites dependent upon $\Delta$ makes little difference (dashed and solid lines of the same colour are similar). Fig.~\ref{fig:hvm_mcf_cV} displays analogous results for the velocity-defined concentration parameter $c_{\rm V}$, suggesting that these results are not driven by any subtle effect of fitting to an NFW density profile. 

These results suggest that much of concentration-dependent clustering at relatively low mass ($M_{200{\rm m}} \lesssim 4 \times 10^{12}\, \hMsun$, masses corresponding to other values of $\Delta$ can be approximated from Fig.~\ref{fig:deltacompare}) is driven by the interactions of nearby haloes. They also suggest that subsuming larger regions into halo definitions to accommodate these interactions may be well motivated and practically useful in the context of halo occupation models. It may even be possible to optimize halo definitions for specific applications. Likewise, these results suggest that at high mass ($M_{200{\rm m}} \gtrsim 4 \times 10^{12}\, \hMsun$), interactions among neighbouring haloes does not appear to be the predominant cause of concentration-dependent clustering. This is consistent with previous literature suggesting that concentration-dependent clustering may be understood as a property of the initial conditions on large-scales according to the excursion set approach \citep{zentner07,dalal_etal08} for high-mass haloes, but is caused by nonlinear interactions at the low-mass end of the halo mass spectrum \citep[e.g.,][]{wang_etal08,warnick_etal08,dalal_etal08,hahn_etal09,ludlow_etal09,lacerna_padilla11,borzyszkowski_etal17}.

Lastly, notice that in the left and middle panels of Fig.~\ref{fig:hvm_mcf_cnfw}, there is some modest residual assembly bias displayed by the solid lines; the solid lines are not all consistent with zero. This suggests that a small part of the reduction in concentration-dependent clustering is caused by the change in the concentration mark values when halo definition changes. Some of the change may come from the introduction of noise in the concentration measurement that is more weakly correlated with large-scale environment upon redefining haloes with larger radii (smaller $\Delta$). Changes in halo masses would also result in changes in the mark values. Nonetheless, for low-mass haloes, this residual assembly bias is generally quite modest compared with the concentration-dependent clustering of haloes defined with more traditional values of the overdensity threshold.

%---- Splashback Comparison
\subsection{Halo Definitions and the Splashback Radius}
\label{subsection:splashback}

Our study is reminscent of, and indeed motivated by, much of the recent work on halo splashback radii \citep{more_etal15, mansfield_etal16, diemer_etal17}. The splashback radius of a halo is the halo-centric distance where infalling material reaches the apocentre of its first orbit. By analogy with the spherical collapse model \citep[e.g.,][]{fillmore_goldreich84}, this radius separates the halo region, within which material is orbiting the halo and we expect interactions to be important, from material on first infall, for which we expect interactions to be less important. The average overdensity enclosed by the splashback radii of haloes is not constant, but varies with halo mass, accretion rate, and redshift \citep{mansfield_etal16, diemer_etal17}. We note that, a priori, there is no reason to expect that using splashback radii and masses would mitigate assembly bias. 

\begin{figure}
	\centering
	\includegraphics[width=\columnwidth]{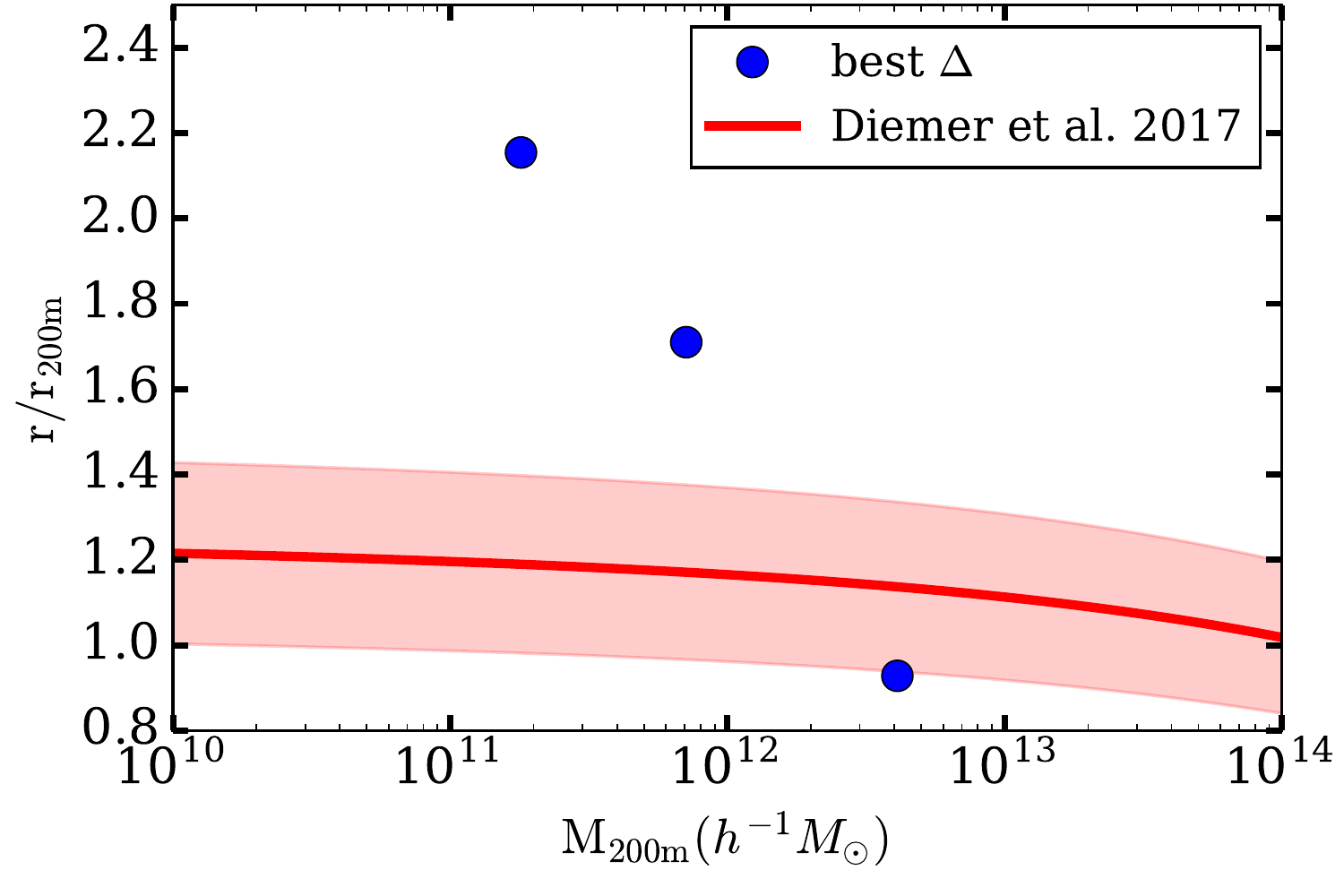}
	\caption{A comparison of the average ratio between $r_{200}$ and the splashback radius as determined by \citet{diemer_etal17} (red line) to the average ratio between $r_{200}$ and the halo radius determined as our best choice of $\Delta$ for removal of assembly bias as discussed above (blue circles). The red shaded region represents the 0.07 dex scatter in the Diemer et al. relation. There is some dispersion in $R_{\Delta}$ but it is quite small (see Fig.~\ref{fig:deltacompare}) so we do not show it on the blue points. Note that the halo mass chosen for the blue points is determined by the mass cutoff in the simulation analysis, as the smallest (and most numerous) haloes dominate the calculation.
}
	\label{fig:splashback_compare}
\end{figure}

Nevertheless, Fig.~\ref{fig:splashback_compare} makes a first connection between our results and splashback radii. The blue points show the radii (in units of $R_{200\text{m}}$) of haloes selected at each of the $\Delta$ thresholds that minimize concentration-dependent halo clustering as a function of halo mass, $M_{200\text{m}}$. Our work suggests that halo definition must be a strong function of halo mass in order to mitigate assembly bias. The red line shows the median splashback radius of haloes as a function of $M_{200\text{m}}$ \citep{diemer_etal17} which exhibits a much weaker dependence on halo mass than needed to mitigate assembly bias. This comparison, however, ignores that the red line is merely an average relation with significant scatter. The overdensities enclosed by the splashback radii of individual haloes do span a range similar to the overdensities found in this paper, where the overdensity is a strong function of the mass accretion rate. Since mass accretion rates are expected to be correlated with large-scale environment, the overall effect of using splashback radii is hard to predict and will be investigated directly in future work.

%-------------- splashback comparison --
%---------------------------------------

%----------------------
\section[]{Conclusions}
\label{section:conclusions}
%----------------------

We have examined the dependence of halo assembly bias upon halo definition, parametrized, for simplicity, by spherical overdensity threshold $\Delta$. Fig.~\ref{fig:halocartoon} presents a pictorial representation of our procedure. This work was motivated as an effort to determine whether or not the dependence of halo clustering strength on halo properties other than mass could be mitigated by judicious choice of halo definition. We summarize our conclusions as follows.

\begin{itemize}
\itemsep0.5\baselineskip

    \item The degree to which halo clustering depends upon auxiliary halo properties varies considerably with halo definition. Even among commonly used halo definitions, such as $\Delta=625$m ($\Delta=200$c), $\Delta=340$m, and $\Delta=200$m, the strength of assembly bias varies considerably with $\Delta$. This is particularly true of halo concentration (see Fig.~\ref{fig:biascompare}). 
    
    \item A judicious definition of a halo can greatly mitigate concentration-dependent halo clustering (see Fig.~\ref{fig:cc_mcf_cnfw} and Fig.~\ref{fig:cc_mcf_cV}). However, this requires a halo definition that has a strong mass dependence; a threshold of $\Delta=20$ mitigates assembly bias at low masses 
($M_{200\text{m}} \sim 1.8 \times 10^{11}\, \hMsun$) but induces assembly bias of opposite sense at high masses ($M_{200\text{m}} \gtrsim 4 \times 10^{12}\, \hMsun$). 
At high masses, an overdensity of $\Delta \approx 340$m results in 
very weak concentration-dependent clustering.

	\item At low mass ($M_{200{\rm m}} \lesssim 4 \times 10^{12}\, \hMsun$ samples from the \simA{} and \simB{} simulations), concentration dependent clustering is mostly driven by low-mass haloes in the immediate neighbourhood of larger haloes that have had their properties altered through interactions with the larger neighbours. This is not the case for more massive haloes (Fig.~\ref{fig:hvm_mcf_cnfw} and Fig.~\ref{fig:hvm_mcf_cV}). Assembly bias at the high-mass end has a distinct origin.
    
    \item Halo shape-dependent clustering is significant over a wide range of halo definitions, but does not exhibit a strong mass dependence for any particular halo definition (Fig.~\ref{fig:cc_mcf_s} and Fig.~\ref{fig:biascompare}). While the trend is for shape dependent halo clustering to be weakened as $\Delta$ is reduced, it cannot be mitigated with any of the halo definitions examined between $20{\rm m} \leq \Delta \leq 625{\rm m}$.
    
    \item Halo spin-dependent clustering demonstrates assembly bias that increases with halo mass. Spin-dependent assembly bias can be mitigated with a threshold of $\Delta \sim 340-625$m for the lowest masses ($M_{200\text{m}} \sim 1.8 \times10^{11}\, \hMsun$), while considerably larger values of $\Delta$ must be used at higher masses ($M_{200\text{m}} \gtrsim 4 \times 10^{12}\, \hMsun$, see Fig.~\ref{fig:cc_mcf_spin} and Fig.~\ref{fig:biascompare}).
    
    \item Although our study was partly motivated by recent studies of the ``splashback radius,'' there is no clear connection between our preferred halo definitions and the average spashback radii of haloes as a function of halo mass (Fig.~\ref{fig:splashback_compare}). However, the splashback radii of individual haloes exhibit scatter about this mean \citep{diemer_etal17} which may be correlated with large-scale environment, meaning that the mean splashback radii may not be a valid representation for such a comparison. A more detailed halo-by-halo exploration is in order.
    
    \item We compile many of our key results in Fig.~\ref{fig:biascompare}, which gives an example of the mass and halo definition ($\Delta$) dependence of the strength of halo assembly bias.
    
\end{itemize}

We conclude that a single, simple, definition of halo size based on overdensity or a similar criterion cannot, by itself, be exploited as a method to mitigate assembly bias. At the very least, mitigation requires any such halo definition to be mass dependent, and likely requires halo definitions that are considerably more complicated than what we have explored here. It is possible that our strategies may be more fruitful when applied to halo properties that we have not studied in the present work. A prominent example of such a property would be a measure of halo formation history. A thorough examination of other potential properties may merit further analysis. Nonetheless, as more and more precise galaxy clustering data become available, we must continue to seek tools that may be used to interpret such high-quality data. Revisiting and reconsidering the concept of a dark matter halo may continue to be one aspect of this search.

\section*{Acknowledgements}

We are grateful to Peter Behroozi, Andreas Berlind, Andrew Hearin, Andrey Kravtsov, Alexie Leauthaud, Surhud More, and Risa \mbox{Wechsler} for a number of informative discussions that influenced our work. Our halo catalogs were generated using {\tt ROCKSTAR} \citep{behroozi_etal13a}. Our calculations were carried out utilizing the {\tt numpy} \citep{numpy}, {\tt astropy} \citep{astropy}, {\tt matplotlib} \citep{matplotlib}, {\tt colossus} \citep{diemer_kravtsov15}, and {\tt halotools} \citep{halotools} packages in Python. The work of ASV and ARZ has been supported, in part, by grants AST 1516266 and AST 1517563 from the U.S.\ National Science Foundation (NSF) as well as by the Pittsburgh Particle physics, Astrophysics, and Cosmology Center (Pitt PACC) at the University of Pittsburgh. YYM is supported by the Samuel P.\ Langley Pitt PACC Postdoctoral Fellowship at the University of Pittsburgh. FCvdB is supported by the Klaus Tschira Foundation and by the US National Science Foundation through grant AST 1516962. BD gratefully acknowledges the support of an Institute for Theory and Computation Fellowship. JUL is supported in part by NSF grant AST 1516962 as well as by a KITP graduate fellowship. KW is supported by NSF grant AST 1517563. 

\bibliographystyle{mnras}
\bibliography{master}

\label{lastpage}

\end{document}

%% file: villarrealzentner_macros.tex
\def\simA{L0125}
\def\simB{L0250}
\def\simC{L0500}

\def\lsim{\lower0.6ex\vbox{\hbox{$ \buildrel{\textstyle <}\over{\sim}\ $}}}
\def\gsim{\lower0.6ex\vbox{\hbox{$ \buildrel{\textstyle >}\over{\sim}\ $}}}

\newcommand{\beq}{\begin{equation}}
\newcommand{\eeq}{\end{equation}}
\newcommand{\beqa}{\begin{eqnarray}}
\newcommand{\eeqa}{\end{eqnarray}}

\newcommand{\Om}{\Omega_{\mathrm{M}}}
\newcommand{\Ol}{\Omega_{\Lambda}}

\newcommand{\Rdel}{r_{\Delta}}

\newcommand{\hMsun}{\ h^{-1}\mathrm{M}_{\odot}~}
\newcommand{\hMpc}{\ h^{-1}\mathrm{Mpc}~}
\newcommand{\hkpc}{\ h^{-1}\mathrm{kpc}~}

\newcommand*{\email}[1]{\href{mailto:#1}{#1}}